\documentclass[%
 aip,
 amsmath,amssymb,
 reprint,%
]{revtex4-1}

\usepackage{graphicx, color}
\usepackage{dcolumn}
\usepackage{bm}
\usepackage{float}

\usepackage[utf8]{inputenc}
\usepackage[T1]{fontenc}
\usepackage{mathptmx}
\usepackage{etoolbox}

\usepackage[dvipsnames]{xcolor}
\usepackage[
    colorlinks=true,
    linkcolor=RoyalBlue,
    citecolor=RoyalBlue,
    urlcolor=RoyalBlue
]{hyperref}

\begin{document}


\title{Mechanically concealed holes}
\author{Kanka Ghosh}
\email{kanka.ghosh@ovgu.de}
\affiliation{ 
Institut f\"ur Physik, Otto-von-Guericke-Universit\"at Magdeburg, Universit\"atsplatz 2, 39106 Magdeburg, Germany
}%
\author{Andreas M. Menzel}%
 \email{a.menzel@ovgu.de}
\affiliation{ 
Institut f\"ur Physik, Otto-von-Guericke-Universit\"at Magdeburg, Universit\"atsplatz 2, 39106 Magdeburg, Germany
}%

\date{\today}

\begin{abstract}
When a hole is introduced into an elastic material, it will usually act to reduce the overall mechanical stiffness. A general ambition is to investigate whether a stiff shell around the hole can act to maintain the overall mechanical properties. We consider this effect from a macroscopic continuum perspective down to atomistic scales. For this purpose, we focus on the basic continuum example situation of an isotropic, homogeneous, linearly elastic material loaded uniformly under compressive plane strain for low concentrations of holes. As we demonstrate, the thickness of the shell can be adjusted in a way to maintain the overall stiffness of the system. We derive a corresponding mathematical expression for the thickness of the shell that conceals the hole. Thus, one can work with given materials to mask the presence of the holes simply by adjusting the thickness of the surrounding shells, with no need to change the materials. Our predictions from linear elasticity continuum theory are extended to atomistic levels using molecular dynamics simulations of a model Lennard-Jones solid. These extensions attest the robustness of our predictions down to atomistic scales. Thus, they open a straightforward possibility to adjust the strategy of mechanical cloaking via atomistic manipulations. From both perspectives, the underlying concept is important in the context of light-weight construction.  
\end{abstract}

\maketitle

\section{Introduction}

Saving resources and fuel is a major concern of recent production lines and construction design. Besides aspects of sustainability, pure economic reasons favor corresponding achievements. Saving materials and energy reduces overall costs \cite{konig2024resource}. Therefore, light-weight construction remains key to future technological developments. Nature provides corresponding examples, maybe bones being the most obvious ones \cite{lai2015dependences,yu2020assessment}. Thanks to their stiff structure they provide overall stability for the whole organism, yet the many cavities of various types of bones reduce their overall weight. This structure saves energy of motion and increases agility and mobility. 

Analogously, for many components of machines, vehicles, aircrafts, or other devices, reducing weight provides significant benefit \cite{koffler2010calculation,timmis2015environmental}. However, the overall design has often been developed for years or decades and been adjusted to near perfection. In such cases, changing the dimension or shape of individual components to reduce their weight, or others of their mechanical properties like stiffness, provides additional challenges. 

Therefore, we focus on the idea of introducing holes into materials to save weight \cite{wang2009structural,zhou2022perforated}. Advanced strategies of designing materials with cylindrical or spherical holes had been already found beneficial for this purpose using approaches of topological optimization \cite{bruggi2022lightweight}. In our case, holes are considered to be mechanically masked in a way so that their presence is not noted on the overall, macroscopic scale of the material. Still, the overall mechanical stiffness shall be maintained. Together, this concept results in a component of identical mechanical properties, yet of reduced weight. Key is to introduce holes (cavities) that are surrounded by stiffer shells so that the overall, combined mechanical stiffness is the same as in the absence of the holes. 

The model geometry of introducing hollow cylindrical shells in three-dimensional solids (or hollow circular rings in two-dimensional solids) has also been applied to study diverse other problems, ranging from elastostatics \cite{lu2017stress, liu2006new} to cavitation in soft solids \cite{kang2018cavitation, kang2021dynamic}.
Amongst these, the idea of mechanically masking holes inside a solid, in fact, has been studied before in terms of ``mechanical cloaking''. It provides the solid with a property of ``mechanical unfeelability" of the holes  within, in terms of the overall macroscopic response. Additionally, mechanically masking holes and corresponding optimization has been realized to conceal mechanical \cite{cheng2023compatible}, elasto-mechanical \cite{buckmann2014elasto, diatta2014controlling, achaoui2020cloaking, stenger2012experiments}, thermo-mechanical \cite{cheng2025thermomechanical}, as well as dispersive properties \cite{meirbekova2020control} of solids, especially in the context of metamaterials design \cite{buckmann2015mechanical, wang2022mechanical, sanders2021optimized, liu2025computationally, craster2023mechanical}. 

Polar materials had been employed as elastostatic cloaks \cite{nassar2018degenerate, zhang2024realizable}, often with functionally graded multilayered lattices \cite{xu2020physical}. Static mechanical cloaking of voids had also been realized recently via a design mechanism of generating irregular structures  \cite{yang2025static}. The paths to identify appropriate  mechanical cloakings is frequently based on advanced transformation methods. They lead to complex structures of the metamaterials design and involve an elevated number of optimized subunits. For instance, the  mechanical cloaking of a hexagonal bimode structure essentially requires stress optimization of the double-trapezoid units in the considered region \cite{hai2018unfeelable}.

Although linear elastodynamic cloaking of cylindrical holes was discussed for infinitesimal in-plane deformations \cite{yavari2019nonlinear}, mechanical concealment in elastostatic situations still receives growing attention. Notably, when viewed from the perspective of optimized materials design, elastostatic cloaking has been outlined following two strategies. Either the displacement fields \cite{fachinotti2018optimization} or the elastic moduli \cite{sozio2023optimal} and thus the type of employed materials were used  as design parameters. We mention that solids containing hollow cylindrical shells, at least concerning mathematical approaches, can be viewed as one representative example of composite materials. Over time, they were studied extensively, with the seminal works of Eshelby marking a notable point of reference \cite{eshelby1957determination}. Others followed \cite{christensen1993effective, Christensen2005}, where studies on composites were concerned mostly with their effective mechanical properties \cite{christensen1979solutions}. Analytical works usually rely on linear elasticity \cite{hsiao1978effective}, and generalized self consistent schemes have been employed \cite{benveniste2008revisiting}. Specific progress on the particle scale considering mutual interactions in analytical theory was achieved when perfectly rigid inclusions were considered \cite{phan1994load,puljiz2019displacement}. Composite structures with holes have also been studied to alleviate stress around holes, albeit using functionally graded materials with varying elastic constants, while keeping Poisson ratios fixed \cite{sburlati2013stress, sburlati2014reduction}. Interestingly, a recent analytical and numerical work \cite{fielding2024simple} showed that elastostatic mechanical cloaking of a circular inclusion in two-dimensional geometries can be attained by coating the inclusion using several concentrically arranged circular rings and tuning their shear moduli. This strategy eliminates the complexity of additionally tuning the Poisson ratio. 

Yet, in reality, the materials to be used may already be determined by other factors, such as constraints of production processes or cost. Prescribing the materials to be used also fixes the elastic moduli, so that these parameters can hardly be adjusted. In that case, a different recipe to implement mechanical shielding and concealing of holes in a given solid is necessary.

Our goal is to present an alternative strategy of mechanical concealment that extends from a macroscopic, continuum level down to microscopic, atomistic scales. To illustrate this concept of mechanically concealed holes, we focus on a basic example situation, keeping the basic geometry as simple as possible. We start from continuum elasticity theory and address uniform static loading under plane-strain conditions of a material that is homogeneous, isotropic, linearly elastic, and, in principle, infinitely extended. The cylindrical holes that we introduce are of sufficiently low concentration so that we may neglect their mutual mechanical interaction.  In reality, we may, for instance, think of cylindrical holes drilled into plates or blocks of material. 
Using analytical calculations, we demonstrate that, indeed, under such conditions, we can introduce stiff cylindrical shells around the cylindrical holes so that the presence of the holes is effectively masked and mechanically concealed. Our sole tuning parameter is the thickness of the shell surrounding the hole. The concept is then extended to atomistic scales using molecular dynamics simulations of an appropriate Lennard-Jones model solid. It maintains valid for discrete particulate structures. We mention that, in reality, colloids often provide example systems to visually bridge between atomistic and macroscopic scales and may serve for corresponding illustrations \cite{herlach2010colloids, royall2024colloidal, menath2023acoustic}.

\section{Continuum elasticity solution} 

\subsection{Theoretical background}

We start our theoretical consideration from the basic theory of linear elasticity \cite{landau2012theory}. Stress $\bm{\sigma}^\mathrm{(3d)}$ and strain $\bm{\varepsilon}^\mathrm{(3d)}$ are related to each other via the shear modulus $\mu$ and the Lam\'e parameter $\lambda$, which is associated with compressibility, 
\begin{equation}\label{eq:stress-strain}
    \bm{\sigma}^\mathrm{(3d)} = 2 \mu\bm{\varepsilon}^\mathrm{(3d)} + \lambda\mathbf{I}\varepsilon_{kk}^\mathrm{(3d)}.
\end{equation}
Einstein summation convention is applied and $\mathbf{I}$ denotes the unit matrix. The second Lam\'e parameter can be expressed by $\mu$ and the Poisson ratio $\nu$ via $\lambda=2\mu\nu/(1-2\nu)$. 

In our plane-strain geometry, we denote in Cartesian coordinates the plane as spanned by coordinates $x$ and $y$, while the direction normal to the plane is referred to by the coordinate $z$. Consequently, plane-strain conditions imply $\varepsilon_{xz}^\mathrm{(3d)}=\varepsilon_{zx}^\mathrm{(3d)}=\varepsilon_{yz}^\mathrm{(3d)}=\varepsilon_{zy}^\mathrm{(3d)}=\varepsilon_{zz}^\mathrm{(3d)}=0$. Thus, solving Eq.~(\ref{eq:stress-strain}) for $\sigma_{zz}^\mathrm{(3d)}$, we can express the whole remaining physics in the two-dimensional plane in terms of the two-dimensional stress $\bm{\sigma}$ and strain $\bm{\varepsilon}$ as
\begin{equation}\label{eq:strain-stress-2d}
    \bm{\varepsilon} = \frac{1}{2 \mu}\left(\bm{\sigma} - \nu\mathbf{I}\sigma_{kk}\right)
\end{equation}
and
\begin{equation}\label{eq:stress-balance}
    \bm{\nabla}\cdot\bm{\sigma}=\bm{0} .
\end{equation}
The latter condition is directly satisfied by deriving $\bm{\sigma}$ from the stress function $F$ as
\begin{equation}\label{eq:stress-F}
    \bm{\sigma} = \mathbf{I}\bm{\nabla}^2 F-\bm{\nabla}\bm{\nabla}F .
\end{equation}

If the associated strain $\bm{\varepsilon}$ derives from a displacement field $\mathbf{u}$, the strain must satisfy certain compatibility conditions. They ensure that, within the framework of linear elasticity, our strain can be expressed as 
\begin{equation}\label{eq:strain-displacement}
\bm{\varepsilon}=\frac{1}{2}\left( \bm{\nabla}\mathbf{u} + \left(\bm{\nabla}\mathbf{u}\right)^T \right) ,
\end{equation}
where $^T$ marks the transpose, independently of whether we start from the Lagrange or Euler framework \cite{chaikin1995principles}. Specifically, in our two-dimensional plane-strain geometry, these compatibility relations reduce to 
\begin{equation}\label{eq:compatibility}
    \nabla_1\nabla_1\varepsilon_{22}+\nabla_2\nabla_2\varepsilon_{11} - 2\nabla_1\nabla_2\varepsilon_{12} = 0  .
\end{equation}
The subscripts $_1$ and $_2$ mark two orthogonal coordinates in the two-dimensional plane. From Eq.~(\ref{eq:strain-stress-2d}) we can calculate the corresponding strain associated with the two-dimensional stress given by Eq.~(\ref{eq:stress-F}). Inserting it into Eq.~(\ref{eq:compatibility}) implies that compatibility with Eq.~(\ref{eq:strain-displacement}) is ensured, if
\begin{equation}\label{eq:biharmonic}
    \bm{\nabla}^2 \bm{\nabla}^2 F=0 .
\end{equation}
Since Eq.~(\ref{eq:stress-balance}) is automatically satisfied in this case, the condition in Eq.~(\ref{eq:biharmonic}) is sufficient and necessary for this solution to exist. 

\subsection{Derivation of the mechanical solution for a shelled hole} 
\begin{figure}
    \centerline{
    \includegraphics[width=7cm]{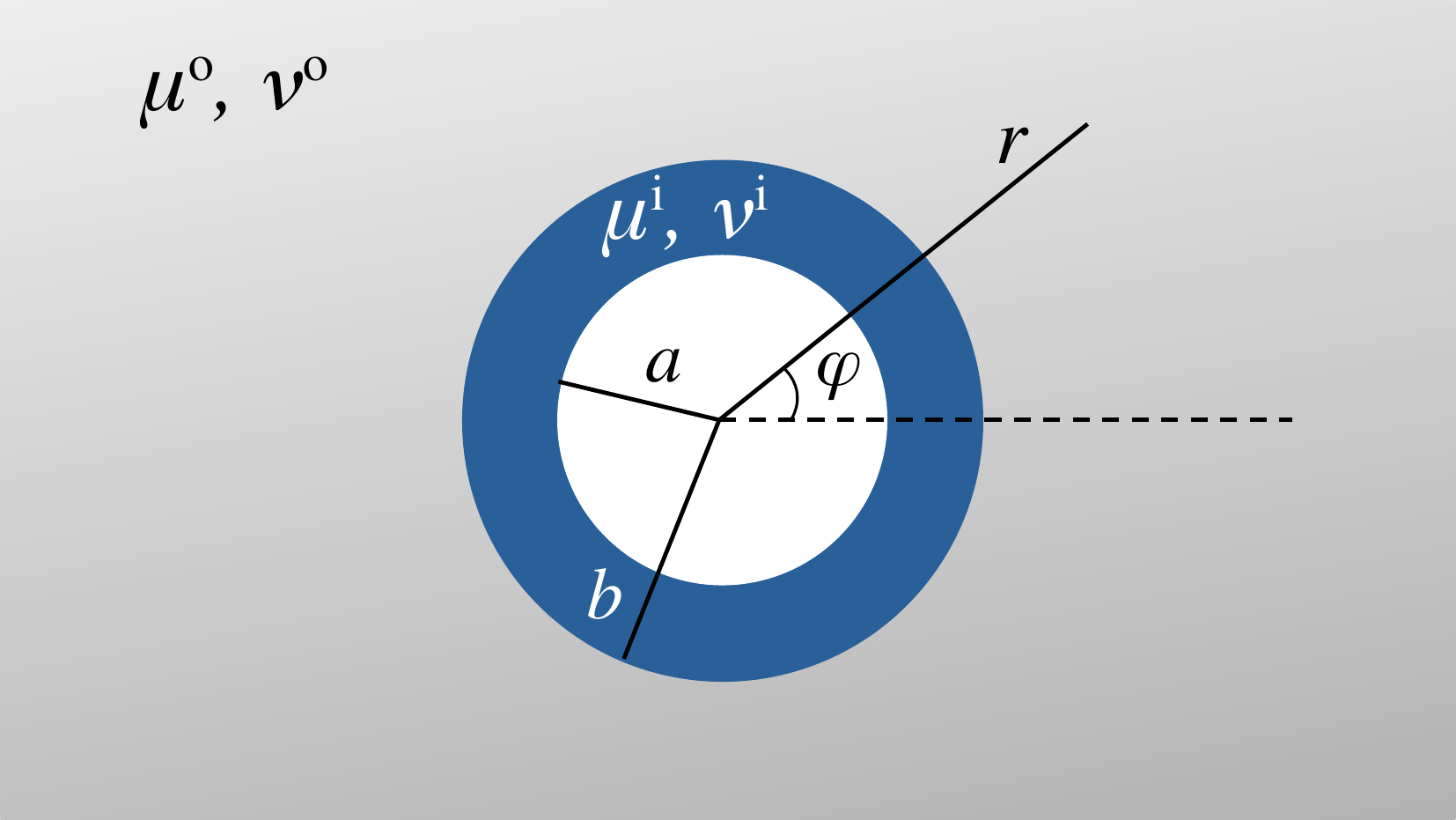}
    }
    \caption{Illustration of the geometry. In the two-dimensional plane that we use to describe the block of material under plane-strain conditions, the cylindrical hole appears as a circular exclusion of radius $a$. Our system of polar coordinates $(r,\varphi)$ is centered in the hole. The hole is surrounded by a cylindrical shell of outer radius $b$, mechanical shear modulus $\mu^\mathrm{i}$, and Poisson ratio $\nu^\mathrm{i}$. Moreover, the actual, outer elastic material is of shear modulus $\mu^\mathrm{o}$ and Poisson ratio $\nu^\mathrm{o}$.}
    \label{fig:geometry}
\end{figure}

We address the situation in polar coordinates. Our coordinate system is centered in the hole, see Fig.~\ref{fig:geometry}. 
The hole has a radius $a$, while the surrounding shell of inner radius $a$ is of outer radius $b>a$. While the shell is of shear modulus $\mu^\mathrm{i}$ and Poisson ratio $\nu^\mathrm{i}$, the corresponding parameters of the surrounding elastic material are $\mu^\mathrm{o}$ and $\nu^\mathrm{o}$. Here and in the following, superscript ``i'' indicates the ``inner'' elastic material (shell), while ``o'' denotes the ``outer'' elastic material. 

\begin{figure*}[t]
   \centering   
    \includegraphics[width=\textwidth]{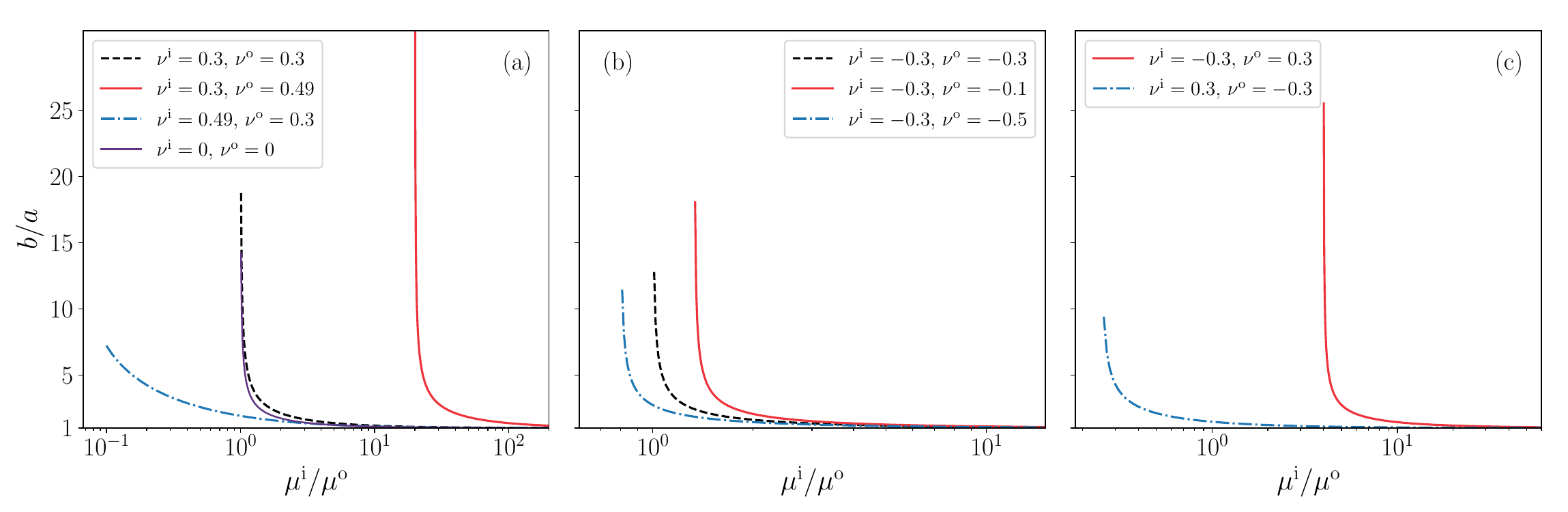}
    \caption{Ratio $b/a$ between the outer radius of the shell $b$ around the hole and the radius $a$ of the hole as a function of the given ratio $\mu^\mathrm{i}/\mu^\mathrm{o}$ between the mechanical moduli of the shell and the surrounding elastic material. The value of $b$ is chosen in a way to mechanically conceal and mask the presence of the hole in the surrounding elastic substance under the imposed deformation. Curves are shown for different combinations of the Poisson ratios $-1<\nu^\mathrm{\{i,o\}}<1/2$ of the shell and the surrounding elastic material, (a) both nonauxetic ($\nu^\mathrm{\{i,o\}} > 0$) , (b) both auxetic ($\nu^\mathrm{\{i,o\}} < 0$), and (c) combined auxetic and nonauxetic ($\nu^\mathrm{i}\nu^\mathrm{o} < 0$).}
    \label{fig:b}
\end{figure*}

Since we consider uniform loading, the stress tensor at infinite distance from the hole can be denoted as $\bm{\sigma}(r\rightarrow\infty)=-P\mathbf{I}$, where $P>0$. In this circularly symmetric situation, for linearly elastic systems, there is no angular dependence of the results on the polar angle $\varphi$. Thus, from all possible terms contributing to the stress function $F$ that satisfy Eq.~(\ref{eq:biharmonic}) \cite{michell1899direct}, we retain only those that do not imply any dependence on $\varphi$ in the physical solution. We formulate the stress function separately for the inner and outer regions, 
\begin{equation}
    F^{\{\mathrm{i},\mathrm{o}\}}=
    A^{\{\mathrm{i},\mathrm{o}\}} r^2
    +B^{\{\mathrm{i},\mathrm{o}\}} \varphi
    +C^{\{\mathrm{i},\mathrm{o}\}} \ln(r).
\end{equation}
In polar coordinates, 
\begin{eqnarray}
    \sigma_{rr}^{\{\mathrm{i},\mathrm{o}\}} &=& \frac{1}{r}\frac{\partial F^{\{\mathrm{i},\mathrm{o}\}}}{\partial r} + \frac{1}{r^2}\frac{\partial^2 F^{\{\mathrm{i},\mathrm{o}\}}}{\partial \varphi^2} , \\
    \sigma_{\varphi\varphi}^{\{\mathrm{i},\mathrm{o}\}} &=& \frac{\partial^2 F^{\{\mathrm{i},\mathrm{o}\}}}{\partial r^2} , \\
    \sigma_{r\varphi}^{\{\mathrm{i},\mathrm{o}\}}&=& \sigma_{\varphi r}^{\{\mathrm{i},\mathrm{o}\}} = {}-\frac{1}{r}\frac{\partial^2 F^{\{\mathrm{i},\mathrm{o}\}}}{\partial r\partial\varphi} + \frac{1}{r^2}\frac{\partial F^{\{\mathrm{i},\mathrm{o}\}}}{\partial \varphi} .
\end{eqnarray}
Thus, Eq.~(\ref{eq:stress-F}) leads us to
\begin{eqnarray}
    \label{eq:sigmarr-A-C}
    \sigma_{rr}^{\{\mathrm{i},\mathrm{o}\}} &=& 2A^{\{\mathrm{i},\mathrm{o}\}} 
    + C^{\{\mathrm{i},\mathrm{o}\}}\frac{1}{r^2} , \\
    \label{eq:sigmaphiphi-A-C}
    \sigma_{\varphi\varphi}^{\{\mathrm{i},\mathrm{o}\}} &=& 2A^{\{\mathrm{i},\mathrm{o}\}} - C^{\{\mathrm{i},\mathrm{o}\}}\frac{1}{r^2} , \\
    \label{eq:sigmarphi-A-C}
    \sigma_{r\varphi}^{\{\mathrm{i},\mathrm{o}\}}&=&\sigma_{\varphi r}^{\{\mathrm{i},\mathrm{o}\}} = B^{\{\mathrm{i},\mathrm{o}\}}\frac{1}{r^2} .
\end{eqnarray}
From Eq.~(\ref{eq:strain-stress-2d}), we find the expressions for the components of the strain tensor
\begin{eqnarray}
    \epsilon_{rr}^{\{\mathrm{i},\mathrm{o}\}} &=& \frac{1}{2\mu^{\{\mathrm{i},\mathrm{o}\}}} \Bigg(  2 \left(1-2\nu^{\{\mathrm{i},\mathrm{o}\}}\right) A^{\{\mathrm{i},\mathrm{o}\}} 
    + C^{\{\mathrm{i},\mathrm{o}\}}\frac{1}{r^2} \Bigg) ,
    \qquad \\
    \epsilon_{\varphi\varphi}^{\{\mathrm{i},\mathrm{o}\}} &=& \frac{1}{2\mu^{\{\mathrm{i},\mathrm{o}\}}} \Bigg(  2 \left(1-2\nu^{\{\mathrm{i},\mathrm{o}\}}\right) A^{\{\mathrm{i},\mathrm{o}\}}- C^{\{\mathrm{i},\mathrm{o}\}}\frac{1}{r^2} \Bigg) , \\
    \label{eq:epsrphi}
    \epsilon_{r\varphi}^{\{\mathrm{i},\mathrm{o}\}}&=& \epsilon_{\varphi r}^{\{\mathrm{i},\mathrm{o}\}} = \frac{1}{2\mu^{\{\mathrm{i},\mathrm{o}\}}}  B^{\{\mathrm{i},\mathrm{o}\}}\frac{1}{r^2} .
\end{eqnarray}
In polar coordinates, the relations between the strain and displacement field 
\begin{eqnarray}
    \epsilon_{rr}^{\{\mathrm{i},\mathrm{o}\}} &=&
    \frac{\partial u_r^{\{\mathrm{i},\mathrm{o}\}}}{\partial r}
     ,\nonumber\\
    \epsilon_{\varphi\varphi}^{\{\mathrm{i},\mathrm{o}\}} &=& 
    \frac{1}{r}\left( \frac{\partial u_\varphi^{\{\mathrm{i},\mathrm{o}\}}}{\partial \varphi} + u_r^{\{\mathrm{i},\mathrm{o}\}} \right)
     ,\nonumber\\
     \label{eq:epsrphi-u}
    \epsilon_{r\varphi}^{\{\mathrm{i},\mathrm{o}\}}&=&\epsilon_{\varphi r}^{\{\mathrm{i},\mathrm{o}\}} =
    \frac{1}{2}\left( \frac{1}{r} \frac{\partial u_r^{\{\mathrm{i},\mathrm{o}\}}}{\partial \varphi} 
    +\frac{\partial u_\varphi^{\{\mathrm{i},\mathrm{o}\}}}{\partial r}
    -\frac{ u_\varphi^{\{\mathrm{i},\mathrm{o}\}}}{r}
    \right)\qquad
\end{eqnarray}
apply. From here, we obtain the displacement fields
\begin{eqnarray}
    u_r^{\{\mathrm{i},\mathrm{o}\}} &=&
    \frac{1}{2\mu^{\{\mathrm{i},\mathrm{o}\}}} \Bigg(  2 \left(1-2\nu^{\{\mathrm{i},\mathrm{o}\}}\right) A^{\{\mathrm{i},\mathrm{o}\}}r 
    - C^{\{\mathrm{i},\mathrm{o}\}}\frac{1}{r} \Bigg) ,
    \qquad \\
    u_\varphi^{\{\mathrm{i},\mathrm{o}\}} &=&{}-
    \frac{1}{2\mu^{\{\mathrm{i},\mathrm{o}\}}} B^{\{\mathrm{i},\mathrm{o}\}}\frac{1}{r}.
    \qquad 
\end{eqnarray}

Our next task is to obtain the values of the coefficients $A^{\{\mathrm{i},\mathrm{o}\}}$, $B^{\{\mathrm{i},\mathrm{o}\}}$, and $C^{\{\mathrm{i},\mathrm{o}\}}$ from the boundary conditions. First, the surface of the hole must be free of traction forces, 
\begin{eqnarray}
    \sigma_{rr}^\mathrm{i}(r=a) &=& 0,
    \\
    \sigma_{r\varphi}^\mathrm{i}(r=a) &=& 0.
\end{eqnarray}
Next, at infinite distance from the hole, the stress must be of the imposed form
\begin{equation}
    \bm{\sigma}^\mathrm{o}(r\rightarrow\infty) = {}-P\mathbf{I}. 
\end{equation}
At the interface between the shell and the surrounding elastic material, both radial stress components must match each other, 
\begin{eqnarray}
    \sigma_{rr}^\mathrm{i}(r=b) &=& \sigma_{rr}^\mathrm{o}(r=b),
    \\
    \sigma_{r\varphi}^\mathrm{i}(r=b) &=& \sigma_{r\varphi}^\mathrm{o}(r=b),
\end{eqnarray}
as must the components of the displacement field, 
\begin{eqnarray}
    u_r^\mathrm{i}(r=b) &=& u_r^\mathrm{o}(r=b),
    \\
    u_\varphi^\mathrm{i}(r=b) &=& u_\varphi^\mathrm{o}(r=b).
\end{eqnarray}

From all these conditions, we find the magnitudes of the coefficients
\begin{eqnarray}
    A^\mathrm{i}&=&
    {}-\frac{\left(1-\nu^\mathrm{o}\right)Pb^2}{\frac{\mu^\mathrm{o}}{\mu^\mathrm{i}}\left[ \left(1-2\nu^\mathrm{i}\right)b^2+a^2\right]+b^2-a^2},
    \\[.1cm]
    A^\mathrm{o}&=&{}-\frac{1}{2}P,
    \\
    \label{eq:B}
    B^\mathrm{i}&=&B^\mathrm{o}={}0,
    \\[.1cm]
    C^\mathrm{i}&=&\frac{2 \left(1-\nu^\mathrm{o}\right)Pa^2b^2}{\frac{\mu^\mathrm{o}}{\mu^\mathrm{i}}\left[ \left(1-2\nu^\mathrm{i}\right)b^2+a^2\right]+b^2-a^2},
    \\
    \label{eq:Co}
    C^\mathrm{o}&=&{}\left(1-\frac{2\left(1-\nu^\mathrm{o}\right)\left(b^2-a^2\right)}{\frac{\mu^\mathrm{o}}{\mu^\mathrm{i}}\left[ \left(1-2\nu^\mathrm{i}\right)b^2+a^2\right]+b^2-a^2}\right)Pb^2.\qquad
\end{eqnarray}
In this way, we have determined the expressions for the displacements, strains, and stresses in the entire domain. For identical materials of the shell around the hole and the surrounding elastic body, that is, for $\mu^\mathrm{i}=\mu^\mathrm{o}$ and $\nu^\mathrm{i}=\nu^\mathrm{o}$, we recover the solution for a uniform elastic body containing a hole, namely $A^\mathrm{i}=A^\mathrm{o}=-P/2$, $B^\mathrm{i}=B^\mathrm{o}=0$, and $C^\mathrm{i}=C^\mathrm{o}=Pa^2$.

\subsection{Thickness of the shell for mechanical concealment} 
We now turn to the central point. The shell around the hole shall mechanically conceal and mask the hole in a way that its presence is not noted from outside. This concealment shall be realized for given materials, that is, we may not modify the material parameters $\mu^{\{\mathrm{i,o}\}}$ and $\nu^{\{\mathrm{i,o}\}}$. For that purpose, we must choose the outer radius $b$ of the shell in a way that the mechanical solution outside the shelled hole appears in the same way as if the hole were not present at all. 

Specifically, this means that the stress outside the shell is given by the imposed uniform stress, 
$\bm{\sigma}^\mathrm{o}(r>b)=-P\mathbf{I}$. From Eqs.~(\ref{eq:sigmarr-A-C})--(\ref{eq:sigmarphi-A-C}), together with Eq.~(\ref{eq:B}), this implies
\begin{equation}
    \label{eq:condition-C}
    C^\mathrm{o}=0.
\end{equation}
This condition further guarantees that also the strain 
$\bm{\varepsilon}^\mathrm{o}(r>b)$ and the displacement field 
$\mathbf{u}^\mathrm{o}(r>b)$ adapt their values in a uniform elastic material as if the hole and its shell were absent.  

Indeed, Eqs.~(\ref{eq:Co}) and (\ref{eq:condition-C}) can be solved for the outer radius $b$ of the shell, 
\begin{equation}
    \label{eq:ratio-b-a}
    b=\sqrt{\frac{\frac{\mu^\mathrm{i}}{\mu^\mathrm{o}}\left(1-2\nu^\mathrm{o}\right)+1}{\frac{\mu^\mathrm{i}}{\mu^\mathrm{o}}\left(1-2\nu^\mathrm{o}\right)-\left(1-2\nu^\mathrm{i}\right)}}\,a.
\end{equation}
The Poisson ratios are confined to values $-1<\nu^{\{\mathrm{i},\mathrm{o}\}}<1/2$. Thus, for shells sufficiently stiffer than the outer elastic material, $\mu^\mathrm{i}\gg\mu^\mathrm{o}$, the expression for $b$ always exists. Complete mechanical shielding of the hole in the surrounding elastic material by a stiff shell is therefore always possible under the given deformation. We illustrate $b$ as a function of the ratio of elastic shear moduli $\mu^\mathrm{i}/\mu^\mathrm{o}$ for different combinations of Poisson ratios $\nu^\mathrm{i}$ and $\nu^\mathrm{o}$ in Fig.~\ref{fig:b}. For identical elastic materials ($\mu^\mathrm{i}=\mu^\mathrm{o}$), whether both of them are nonauxetic as in Fig.~\ref{fig:b}(a) or auxetic as in Fig.~\ref{fig:b}(b), the necessary outer radius of the shell and thus the necessary thickness of the shell $b-a$ diverge. In contrast to that, for very stiff shell materials $\mu^\mathrm{i}\gg\mu^\mathrm{o}$, the thickness $b-a$ of the shell tends towards zero ($b/a\rightarrow 1$ in Fig.~\ref{fig:b}). 

If the Poisson ratio of the shell is smaller compared to the surrounding material, $\nu^\mathrm{i}<\nu^\mathrm{o}$, the necessary outer radius and thus thickness of the shell diverges even with shells stiffer than the outside elastic material, $\mu^\mathrm{i}>\mu^\mathrm{o}$, if the shells are not overly stiff, see the red curves in Fig.~\ref{fig:b}. In such cases, for given materials, mechanical shielding will be challenging, and a stiffer choice of shell material may still be advisable. For example, the combination $\nu^\mathrm{i} = 0.3$ and $ \nu^\mathrm{o} = 0.49$, see Fig.~\ref{fig:b}(a), depicts a compressible shell such as steel \cite{greaves2011poisson, mott2009limits} within a nearly incompressible background like a rubbery elastomer \cite{greaves2011poisson, mott2009limits}. However, given that steels are generally several orders of magnitude stiffer than elastomers, this does not represent an actual constraint for mechanical concealment. It would still be possible already using very thin shells. 

Interestingly, mechanical concealment would also be possible in the reverse situation of $\nu^\mathrm{i} > \nu^\mathrm{o}$ for a hard incompressible elastomeric shell surrounding a hole in a soft compressible solid, as long as the shell is not substantially softer than the surrounding solid, $\mu^\mathrm{i}\lessapprox\mu^\mathrm{o}$. This statement applies irrespectively of the absolute values of $\nu^\mathrm{\{i,o\}}$, see the blue dash-dotted lines in Fig.~\ref{fig:b}. Yet, mechanical concealment of the hole is only achieved at elevated thicknesses of the shell ($5 < b/a < 15$).

Thus, where possible, selecting most beneficial materials provides advantages. Leveraging the existence of various unconventional materials, one may even opt for materials of near-zero or negative Poisson ratios \cite{soman2012three, jiang2014negative, yu2017negative, liu2018soft, lee2019auxetic, gaal2020new, gao2024strain} to achieve advanced mechanical shielding. For instance, graphene monolayers can be rendered auxetic by including holes and patterned defects \cite{grima2015tailoring, grima2018giant}. However, once a choice is made or the materials are determined by other factors, then still for many practical material combinations the strategy of simply adjusting the thickness of the shell provides an effective, realizable way for mechanical concealment. 
We recall that we assumed uniform imposed deformations in the presence of one hole. Confining ourselves to linear elasticity theory, the solution for several holes can simply be superimposed. However, we neglect in this framework their mutual interactions by surrounding deformations induced by the holes. Therefore, our approach is currently confined to low densities of noninteracting holes to remain quantitative.

\section{Molecular dynamics simulations of mechanical cloaking} 
To proceed a step beyond continuum considerations, we seek to investigate whether our analytical solution for mechanical concealment of a hole carries over to atomistic levels. If consistency between continuum and atomistic solutions is identified, it will indicate a robust cloaking strategy for large-scale applications. Thus, a scope of microscopic realization and adjustment of atomistic mechanical properties is conceivable. This correlation would allow us to connect continuum elasticity with microscopic structural stability of matter in the context of cloaking. To this end, we analyze a microscopic model solid using molecular dynamics (MD) simulations. For given combinations of $\mu^\mathrm{i}/\mu^\mathrm{o}$ and $b/a$, we test whether mechanical concealment can be achieved. 

In fact, atomistic investigations of mechanically concealed holes inside a solid are rare and challenging. Specifically, stabilizing solids with voids or cavities on the atomic scale often needs many-body terms in the interatomic interaction potentials. They come at a cost of losing simple, transparent models. Often, the descriptions invoke anisotropy due to angular dependence of the potentials. Our scope is to circumvent this issue and work with a description as simple as possible. Therefore, we use a two-dimensional, truncated and shifted Lennard-Jones (LJ) potential and adjust its length and energy scale, that is, the depth of the energy well, see Appendix \ref{app1} for details, even if resulting energy wells are significantly deeper than for standard rare-gas solids. They imply very high mechanical moduli of the order of TPa. As an advantage, we can work with a very simple and accessible potential and thus mimic a ``toy model'' for a strongly cohesive solid. As a further benefit, the LJ potential in two dimensions leads to stable hexagonal solids. The linearly elastic properties of solids of hexagonal symmetry are the same as for isotropic ones from a continuum perspective \cite{landau2012theory}. Therefore, our atomistic approach is in line with the continuum considerations above for isotropic elastic materials.

\subsection{Simulation details and protocols \label{sec:md1}}

Molecular dynamics (MD) simulations are carried out in two dimensions using 46200 atoms interacting via a truncated and force-shifted Lennard-Jones (LJ) potential, see Fig.~\ref{fig:lj} in Appendix. It is defined as \citep{toxvaerd2011communication}

\begin{equation}
\begin{split}
V_{SF}(r) =
\begin{cases}
V_{LJ}(r) - V_{LJ}(r_{c}) \\
\quad - (r - r_{c}) V_{LJ}^{\prime}(r_{c}), & r < r_c, \\[6pt]
0, & r \ge r_c,
\label{eq:lj1}
\end{cases}
\end{split}
\end{equation}
with
\begin{equation}
    V_{LJ}(r) = 4\epsilon\left[\left(\frac{\sigma}{r}\right)^{12} - \left(\frac{\sigma}{r}\right)^{6} \right],
\label{eq:lj2}
\end{equation}
where $V_{SF}(r)$ and $V_{LJ}(r)$ are the force-shifted and the standard Lennard-Jones interatomic potentials, respectively. Here, $r$ denotes the center-to-center distance between two atoms. $r_{c}$ denotes the cutoff distance of the potential ($=2.5$~\AA{}). To mimic a ``toy model'' of a strongly cohesive two-dimensional background solid, $\epsilon=\epsilon^\mathrm{o}$ = 1.0 eV and $\sigma$ = 1.0~\AA{} have been chosen as energy and length scale parameters respectively. Within the shell surrounding the hole, we set $\epsilon=\epsilon^\mathrm{i}$. The Lorentz-Berthelot mixing rule ($\epsilon = \sqrt{\epsilon^{\mathrm{i}}\epsilon^{\mathrm{o}}}$) is used to define the energy parameter of the potential between an atom belonging to the shell and an atom being part of the surrounded elastic solid. $\sigma$ = 1.0~\AA{} is considered for all atoms in the solid. 

\begin{figure}[t]
\centering
 \includegraphics[width=7cm]{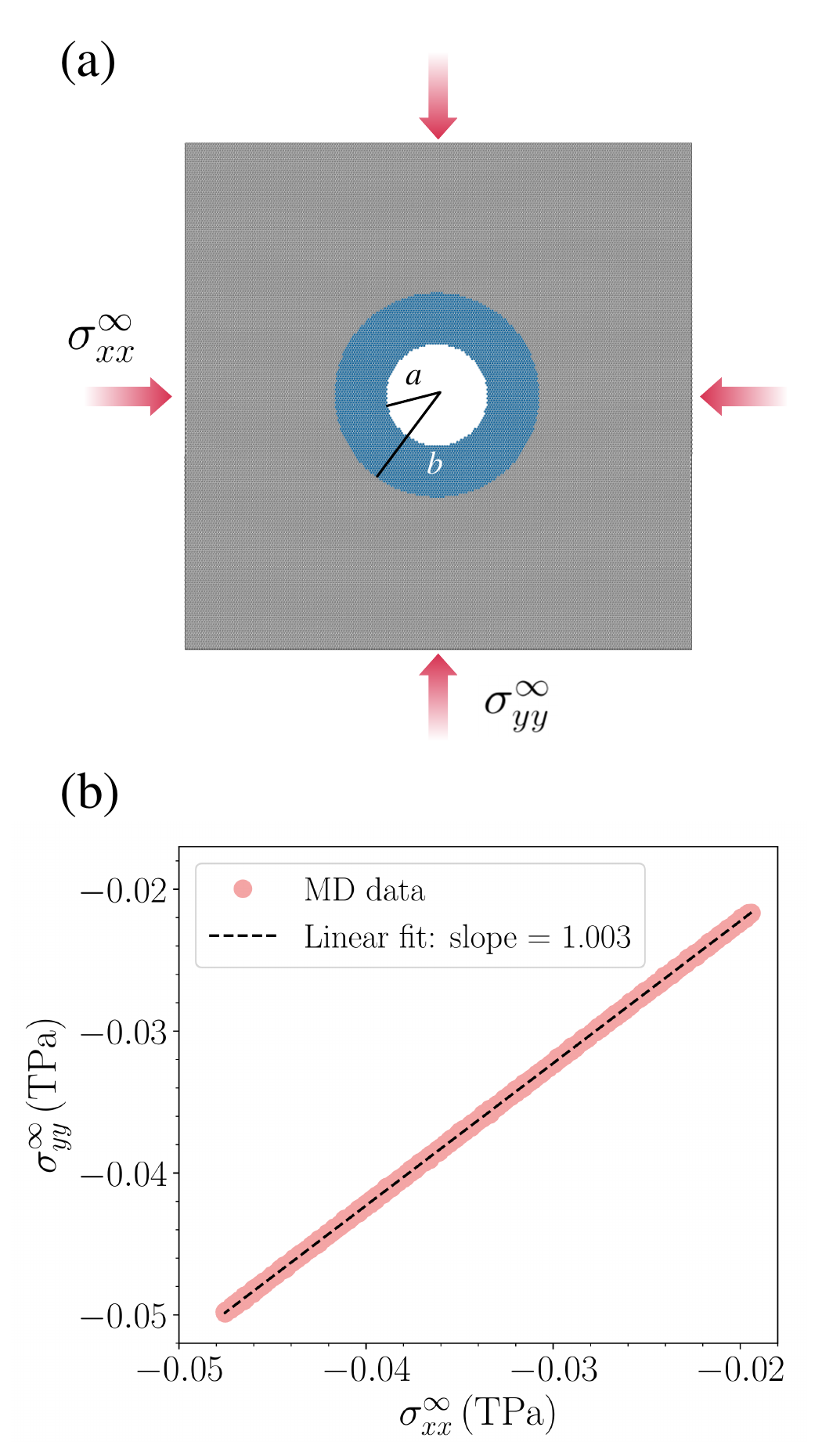}
 \caption{\label{fig:iso_comp} (a) Illustration of a representative isotropic compression test in an MD simulation of a planar hexagonal solid with a shielded hole ($b/a = 2.018$, $\mu^\mathrm{i}/\mu^\mathrm{o} = 2$). Isotropic compression can be realized by imposing an isotropic stress $\bm{\sigma}^\infty=\bm{\sigma}(r\rightarrow\infty)=-P\mathbf{I}$, see the main text. (b) Isotropic compression is confirmed by plotting the components of the imposed stress $\sigma^\infty_{xx}$ and $\sigma^\infty_{yy}$ against each other. The linear fit confirms an approximate 
 unit slope.}
\end{figure}

We impose periodic boundary conditions (PBC) along both $x$ and $y$ directions. A very low temperature ($T$ = 0.6 K) is maintained throughout the simulations to minimize the thermal effects and facilitate direct comparison with the continuum theory. We equilibrate systems using isothermal-isobaric (NPT) ensemble for 10$^6$ steps with a time step $\Delta t$ of 0.0001~ps at a considerably higher pressure of 2$\times$10$^{5}$~bar using velocity Verlet algorithm. At this pressure, samples equilibrate with dimensions $L_{x} = L_{y} = L \approx 223$~\AA{}. The higher pressure stabilizes the planar solid, regardless of the presence of the hole. 

For hexagonal solids with holes, mechanical concealment is achieved by introducing a concentric circular ring-like region (shell) of variable thickness and variable stiffness around the hole. A Lennard-Jones potential to realize our model solid is particularly useful in this context. In this case, we can directly relate the stiffness in terms of the shear modulus $\mu$ to the LJ energy parameter $\epsilon$ via $\mu$ $\approx$ ${\epsilon}/{r_{0}^2}$, with $r_{0}$ denoting the lattice constant (see Section~\ref{sec:md2} and Fig.~\ref{fig:rdf} in Appendix for details). Further, these pristine as well as hexagonal solids with holes (with and without shielding shell around the hole) are subjected to isotropic compression tests. 

Holes were carved out with a radius $a$ of 20~\AA{}. To this end, computational isotropic compression tests are performed on the equilibrated samples via unbiased area contraction using a constant engineering strain rate of $10^{-4}$ per timestep. We run these simulations for $5\times10^6$ MD steps (corresponding to 0.5~ns) within NVT ensemble, out of which only the initial small-strain, linear pressure-areal strain regimes (up to 0.3 $\%$ areal strain) are used to compute the bulk modulus $K$ in two dimensions via

\begin{equation}\label{eq:bulk}
    K = - A_{0}\left( \frac{\partial P}{\partial A} \right)_{T} = - \left(\frac{\partial P}{\partial \epsilon_{A}}\right)_{T}, 
\end{equation}
where we defined the areal strain $\epsilon_{A} = {\Delta A}/{A_{0}}$. Here, $A_{0}$ denotes the equilibrated area 
and $\Delta A$ represents the change in area from the reference area $A_{0}$ during isotropic compression to the new area $A$.

In Eq.~(\ref{eq:bulk}), the negative sign signifies decreasing area under compression (increasing pressure). The setting for a typical isotropic compression test is illustrated in Fig.~\ref{fig:iso_comp}(a) for an equilibrated system with a shielded hole. For the same sample, isotropic compression can be confirmed from the unit slope between the $xx$- and $yy$-components of the imposed stress, see Fig.~\ref{fig:iso_comp}(b).

Mechanical concealment of holes is finally achieved using the following protocol. First, we choose a given stiffness of the shells surrounding the holes using a specific ratio of shear moduli $\mu^\mathrm{i}/\mu^\mathrm{o}$. Then, we tune the thickness of the shell by suitably varying the ratio between the outer radii of shell and hole $b/a$. The concealment ratio is the desired $b/a$ that gives identical bulk modulus ($K$) of the solid with hole to that of the pristine solid without any hole (within a relative error of $\leq 0.07 \%$). 
Throughout our MD simulations we consider $\nu^\mathrm{i}$ $\approx$ $\nu^\mathrm{o}$ = 0.34 (see Section~\ref{sec:md3} for details). All molecular dynamics simulations described in this article are performed using \textsc{LAMMPS} \citep{thompson2022lammps, Lammps}. 

\begin{figure}[!htb]
\centering
 \includegraphics[width=7cm]{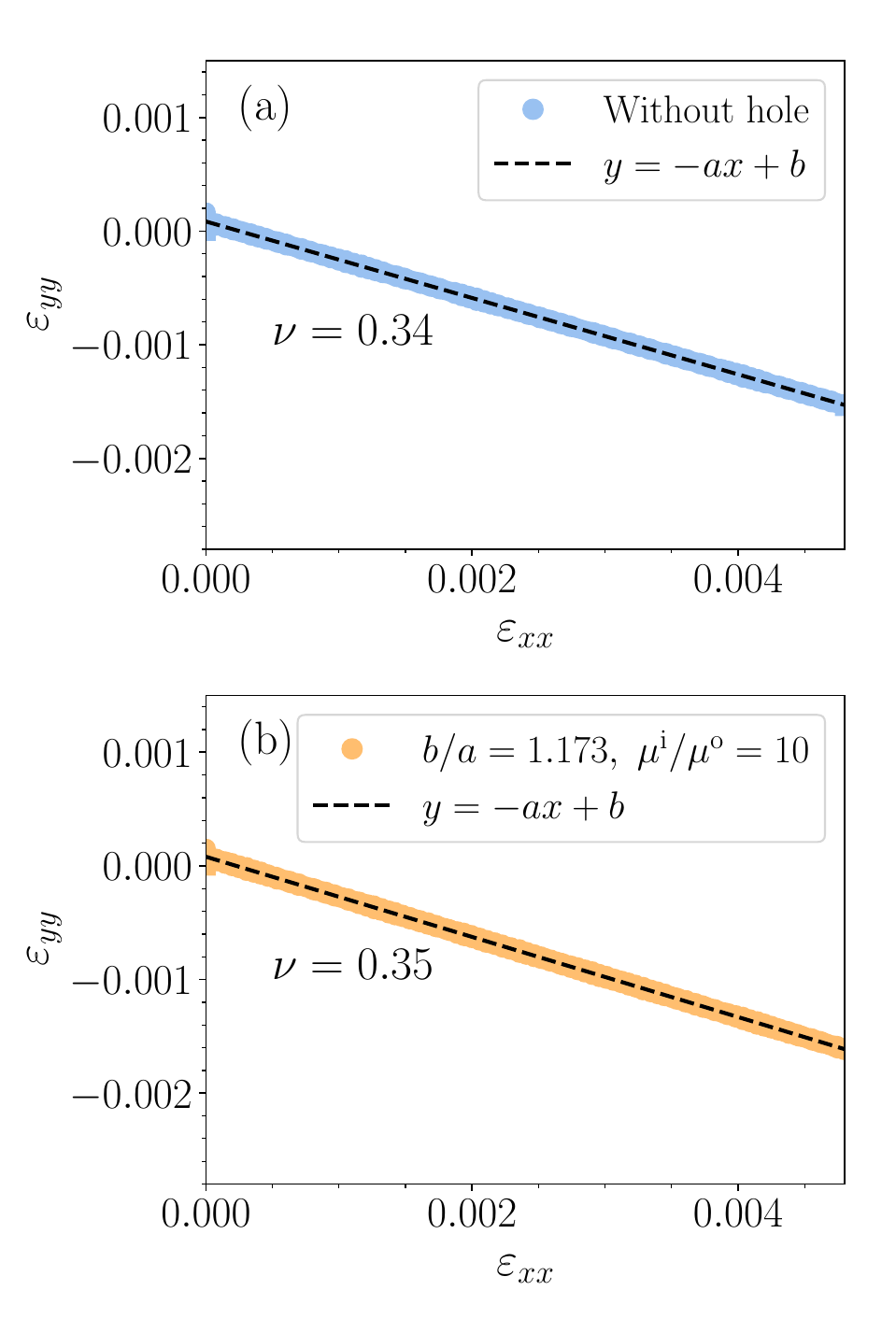}
 \caption{\label{fig:poisson} Poisson ratios ($\nu$), obtained from the slope of the linear variation between an applied strain ($\epsilon_{xx}$) and a resulting transverse strain ($\epsilon_{yy}$) via performing uniaxial tensile tests of the equilibrated samples. Results are shown for (a) a pristine planar solid and (b) a planar solid with a hole enclosed by a stiffer shell of ratios of outer radii $b/a = 1.173$ and of shear moduli $\mu^\mathrm{i}/\mu^\mathrm{o} = 10$.} 
\end{figure}

\subsection{Controlling the shear modulus of the shell \label{sec:md2}}

According to Ref.~\citep{fisher1979defects}, the elastic moduli of an isotropic, homogeneous lattice with two Lam\'e coefficients $\mu$ and $\lambda$ under short-range potentials can be obtained from the long-wavelength dispersion relations via
\begin{equation}\label{eq:omeg_t}
    m\omega_{T}^{2} = \mu (qr_{0})^{2},
\end{equation}
\begin{equation}\label{eq:omeg_l}
    m\omega_{L}^{2} = (\lambda+2\mu) (qr_{0})^{2}.
\end{equation}
Here, $\omega_{L,T}$ are the longitudinal and transverse phonon frequencies, $q$ denotes the wave number, $m$ represents the effective mass, 
and $r_0$ is the equilibrium lattice parameter. 

As shown in Ref.~\citep{zhang2021phonon}, for particles in a hexagonal lattice interacting via LJ potentials, long-wavelength dispersion curves yield  
\begin{equation}\label{eq:omeg_t2}
    m\omega_{T}^{2} = \epsilon \frac{27}{r_{0}^{2}} (qr_{0})^{2},
\end{equation}
\begin{equation}\label{eq:omeg_l2}
    m\omega_{L}^{2} = \epsilon \frac{27\times3}{r_{0}^{2}} (qr_{0})^{2}. 
\end{equation}

Therefore, comparing Eqs.~(\ref{eq:omeg_t})--(\ref{eq:omeg_l2}), we find
\begin{equation}
    \lambda = \mu = 27\frac{\epsilon}{r_{0}^{2}}.
\end{equation}
Here, $\epsilon$ is the LJ energy parameter. $r_{0}$ is fixed to 1.12~\AA{} in our simulations, see Fig.~\ref{fig:rdf} in Appendix. Thus, the relation between the ratios of shear moduli and the LJ energy parameters for shell and surrounding solid becomes $\mu^\mathrm{i}/\mu^\mathrm{o} = \epsilon^\mathrm{i}/\epsilon^\mathrm{o}$.

\begin{figure}[b]
    \centerline{
    \includegraphics[width=9cm]{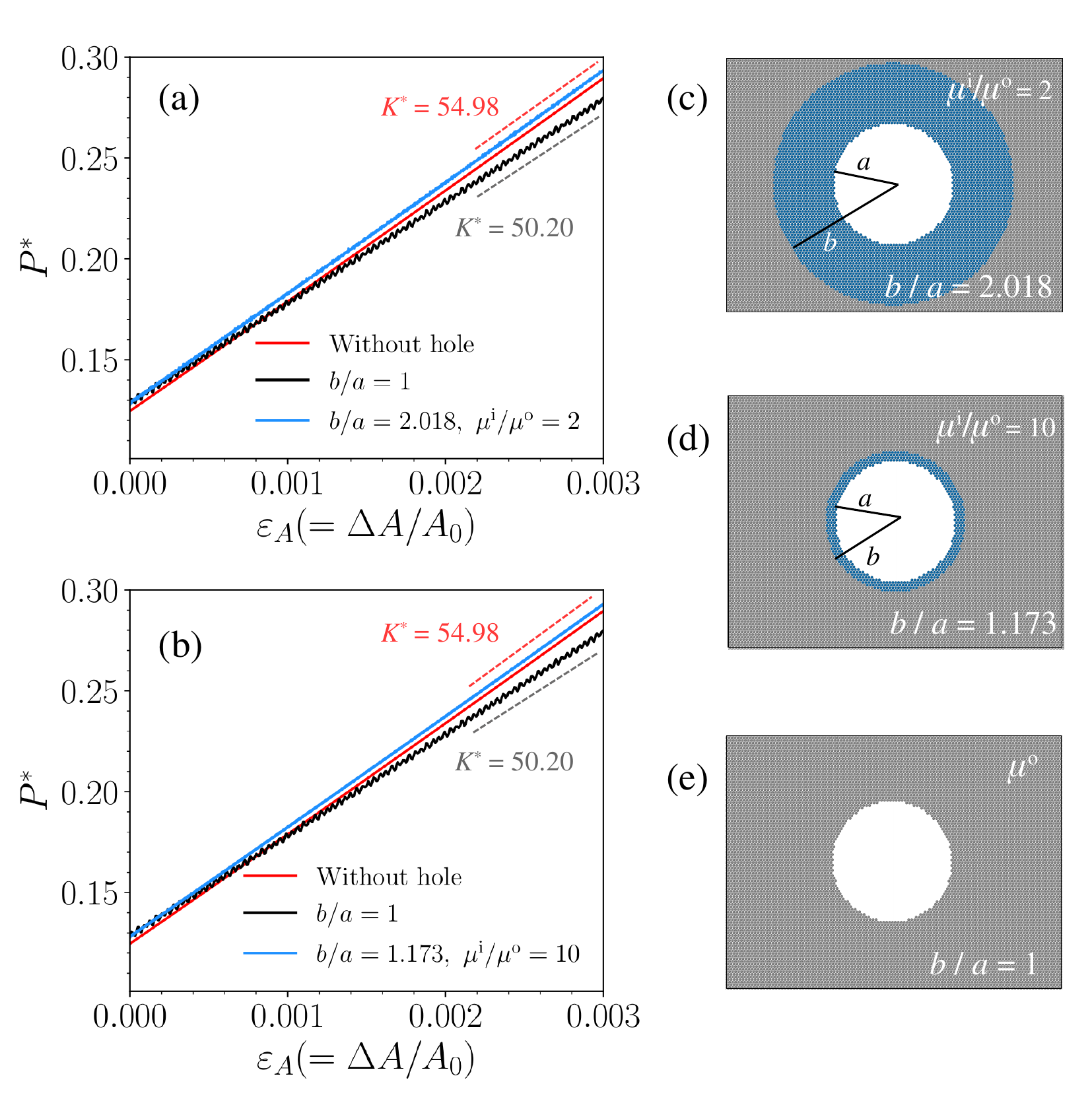}
    }
    \caption{MD simulation results for the variation of pressure $P^*$ with areal strain $\epsilon_{A} = \Delta A/A_{0}$ during isotropic compression. Here, $A_0$ is the area of the equilibrated system at a given initial pressure $P^*(\epsilon_A=0)$, and $\Delta A$ is the change in area from there. We define the bulk moduli $K^*$ as the slopes of the resulting curves. Both pressure $P^*=P\sigma^{2}/\epsilon$ and bulk modulus $K^*=K\sigma^{2}/\epsilon$ are rescaled by the LJ parameters.  
    We consider the two ratios of elastic moduli between the shell and the surrounding solid (a) $\mu^\mathrm{i}/\mu^\mathrm{o} = 2$ and (b) $\mu^\mathrm{i}/\mu^\mathrm{o} = 10$. 
    In both cases, $P^*(\epsilon_A)$ is plotted for the pristine planar hexagonal solid without any hole (red curves), the solid with an unshelled hole ($b/a=1$, black curves), and for the solid with a shelled hole of a thickness $b/a>1$ (blue curves) that best ensure mechanical concealment (same slope $K^*\approx54.98$ for red and blue curves). The corresponding geometries were identified as (c) $b/a = 2.018$ and (d) $b/a=1.173$, respectively, in contrast to (e) the unshelled hole that yields a lower $K^*\approx50.2$. The side panels show simulation snapshots with gray and blue atoms denoting whether they belong to the background solid or shell, respectively. We considered $\nu^\mathrm{i}$ $\approx$ $\nu^\mathrm{o}$ $= 0.34$ in all cases.}
    \label{fig:md_1}
\end{figure}

\begin{figure*}[!htb]
   \centering   
    \includegraphics[width=\textwidth]{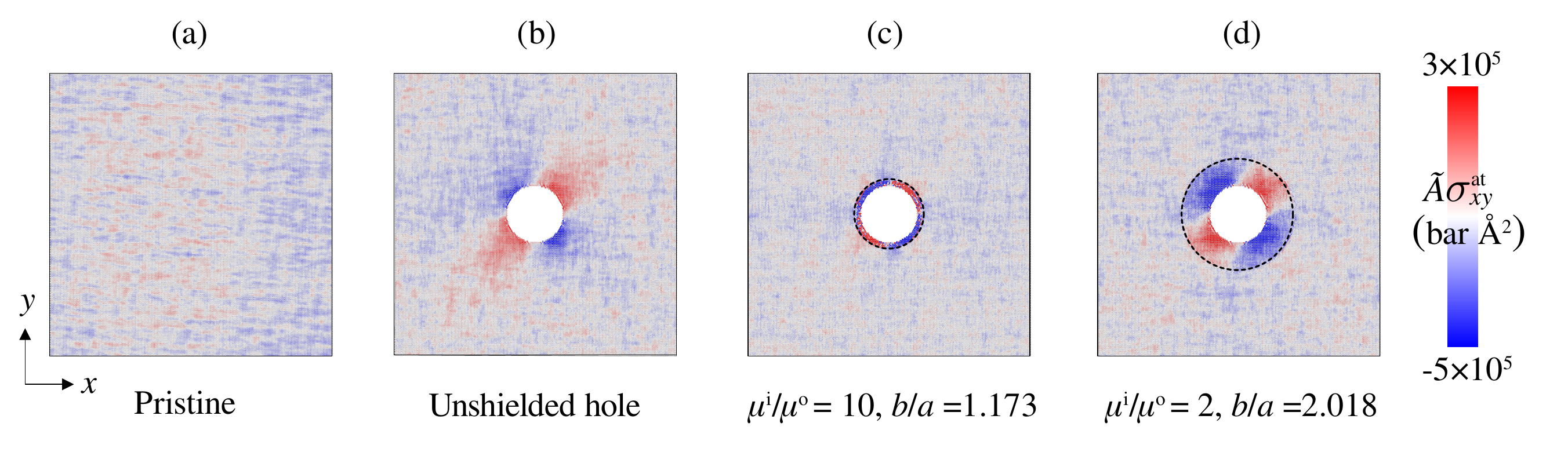}
    \caption{Atomistic distributions of instantaneous virial shear stress during isotropic compression at an areal strain of 0.3~$\%$. We consider (a) a pristine elastic solid without any hole, (b) an elastic solid with an unshielded hole ($b/a$ = 1), (c) an elastic solid with a shielded hole of parameters $b/a = 1.173$,  $\mu^\mathrm{i}/\mu^\mathrm{o} = 10$, and (d) an elastic solid with a shielded hole and parameters $b/a = 2.018$,  $\mu^\mathrm{i}/\mu^\mathrm{o} = 2$. The outer perimeter of the shielding shell of radius $b$ is shown via a black, dashed circle in (c) and (d). \textsc{OVITO} \cite{stukowski2009visualization} is used for visualization.}
    \label{fig:stress}
\end{figure*}

\subsection{Poisson ratio evaluation \label{sec:md3}}

In a hexagonal lattice composed of atoms interacting via an LJ potential, we noted for the Lam\'e coefficients $\lambda = \mu$. Therefore, the Poisson ratio becomes $\nu = \lambda/(2\mu +\lambda) = 1/3$. Indeed, from our MD simulations, we extract $\nu = 0.34$ from the slope of the linear variation between transverse strain ($\epsilon_{yy}$) and applied strain ($\epsilon_{xx}$) via uniaxial tensile test of the equilibrated pristine hexagonal solid, see Fig.~\ref{fig:poisson}(a). LJ solids in two dimensions can exhibit a wide range of values of the Poisson ratio $\nu$, depending on the applied pressure or the presence of point defects \cite{wojciechowski1999computer}. They can even feature auxetic behavior, depending on the lattice constants and at negative pressure \cite{rechtsman2008negative}. However, as shown in Fig.~\ref{fig:poisson}(b), introducing a stiff shell surrounding a hole does not significantly alter $\nu$ in our case, here to a value of approximately 0.35. Throughout the evaluations of our MD simulations, we thus assume $\nu^\mathrm{i}$ $\approx$ $\nu^\mathrm{o}$ =  0.34. 

We note here that the Poisson ratio in two-dimensional hexagonal systems experiences an upper limit of 1 \cite{kw2003remarks, gao2021bounds}, compared to the three-dimensional case, where it is bound by $1/2$. Our obtained Poisson ratio for the two-dimensional LJ solid was evaluated for a genuinely two-dimensional system. It is approximately 0.34, so we are well within the given limits. Comparison between the results suggested by continuum elasticity theory for the considered range of thermodynamically stable Poisson ratios under plane-strain conditions and the selected atomistic two-dimensional LJ model solid is therefore possible. The Poisson ratios listed in the expressions derived from continuum elasticity theory, Eq.~(\ref{eq:ratio-b-a}), are still the three-dimensional ones. Therefore, to compare our MD results with continuum theory, we employ the conversion formula for plane strain \cite{eischen1993determining, meille2001linear}, $\nu^{\{i,o\}}=\nu^{\{i,o\},MD}/(1+\nu^{\{i,o\},MD})$. In this expression, $\nu^{\{i,o\},MD}$ is our value of approximately $0.34$ obtained from MD simulations, which converts to approximately $\nu^{\{i,o\}}\approx 0.254$ for comparison with the result in Eq.~(\ref{eq:ratio-b-a}). The shear moduli remain identical under this conversion \cite{eischen1993determining}.

\subsection{MD results: shell thickness for concealment \label{sec:md4}}

Pristine hexagonal solids, hexagonal solids with a hole but no shell ($b/a=1$), and hexagonal solids containing shelled holes ($b/a>1$) were prepared. We here use the bulk modulus to compare their mechanical behavior and test the effective mechanical shielding of the hole from isotropic compression. 

We proceed as follows to achieve mechanical concealment of the hole. 
Initially, we choose a given stiffness of the shell surrounding the hole by setting a specific value of the ratio $\mu^\mathrm{i}/\mu^\mathrm{o}$. This step corresponds to accepting the situation of given materials that we need to work with. Then we tune the thickness of the shell by suitably varying $b/a$. The desired $b/a$ is obtained by matching the bulk modulus $K$ of the hexagonal solid with shelled hole to that of the pristine planar hexagonal solid without hole. 
Figure~\ref{fig:md_1} represents two examples of such numerical experiments, both for $\nu^\mathrm{i}$ $\approx$ $\nu^\mathrm{o}$ = 0.34, yet for (a) $\mu^\mathrm{i}/\mu^\mathrm{o}=2$ and (b) $\mu^\mathrm{i}/\mu^\mathrm{o}=10$. 
In both Figs.~\ref{fig:md_1}(a) and (b), the pristine hexagonal solid without any hole yields an elastic bulk modulus $K=54.98\epsilon/\sigma^2$ (red curves). When we cut out the hole, but do not put any shell surrounding it ($b/a=1$), see Fig.~\ref{fig:md_1}(e), the bulk modulus is reduced to $K=50.2\epsilon/\sigma^2$ (black curves). However, we can restore the initial bulk modulus of $K=54.98\epsilon/\sigma^2$ by placing a stiff shell of matched thickness enclosing the hole (blue curves). The corresponding thicknesses to conceal the presence of the hole are given by $b/a\approx2.018$ for the softer shell material $\mu^\mathrm{i}/\mu^\mathrm{o} = 2$, see Fig.~\ref{fig:md_1}(c), and $b/a\approx1.173$ for the stiffer shell material $\mu^\mathrm{i}/\mu^\mathrm{o} = 10$, see Fig.~\ref{fig:md_1}(d). The procedure demonstrates that effective concealment of the hole works also on this atomistic scale. As expected, the necessary thickness of the shell is larger for the softer shell material. 

Moreover, we investigate the effect of mechanical shielding of holes on the stress distribution of the solid on the atomistic level. We recall that the shielding strategy relies on tuning the ratio $b/a$, so that the effective bulk modulus matches that of the pristine solid. Thereby, we account for the stress components $\sigma_{xx}$ and $\sigma_{yy}$. Ideally, macroscopic isotropic compression does not intend to impose global shear stresses. Still, fluctuations on the atomic level persist and could give rise to microscopic, instantaneous virial shear stresses around the hole. This naturally raises the more specific, subsequent question about whether the cloak mitigates these instantaneous virial shear stresses during the compressive response at the atomistic level.

Figure~\ref{fig:stress} represents the distribution of virial shear stress on the atoms during isotropic compression for pristine, unshielded as well as for concealed holes. The areal strain is set to 0.3~$\%$ to remain within the elastic limit. As a measure for the atomic shear stress, 
we follow the expression \cite{thompson2009general}
\begin{equation}\label{eq:virial}
    \tilde A\sigma_{xy}^{\text{at}} =  \sum_{i=1}^{N} \left[{}-m_{i}v_{i,x}v_{i,y} - \sum _{j>i} r_{ij,x} F_{ij,y}  \right].
\end{equation}
Here, $\tilde A$ is a typical area of the order of the area of one atom. $m_i$ is the mass of particle $i$. $x$- and $y$-components of the velocity of particle $i$ are defined as $v_{i,x}$  and $v_{i,y}$, respectively. $r_{ij,x}$ denotes the $x$-component of the distance and $F_{ij,y}$ describes the $y$-component of force between particles $i$ and $j$. Since the kinetic term is typically very small, we neglect it in the calculation of the shear stress. 

The pristine solid without the hole in Fig.~\ref{fig:stress}(a) shows a rather scattered distribution of local stress. Contrarily, instantaneous shear stresses are observed to build up and accumulate around an unshielded hole, see Fig.~\ref{fig:stress}(b). They reach into the surrounding solid to a notable distance. Upon cloaking the hole, that is, inserting the stiffer shell, we observe that the desired $b/a$ for concealment effectively confines the instantaneous virial shear stresses around the hole to the region within the shell thickness, see Fig.~\ref{fig:stress}(c,d). The background solid remains in these cases with similar scattered distributions of virial shear stress as for the pristine solid without the hole in Fig.~\ref{fig:stress}(a).

\subsection{Comparing atomistic with continuum results}

\begin{figure}[t]
    \centerline{
    \includegraphics[width=8.5cm]{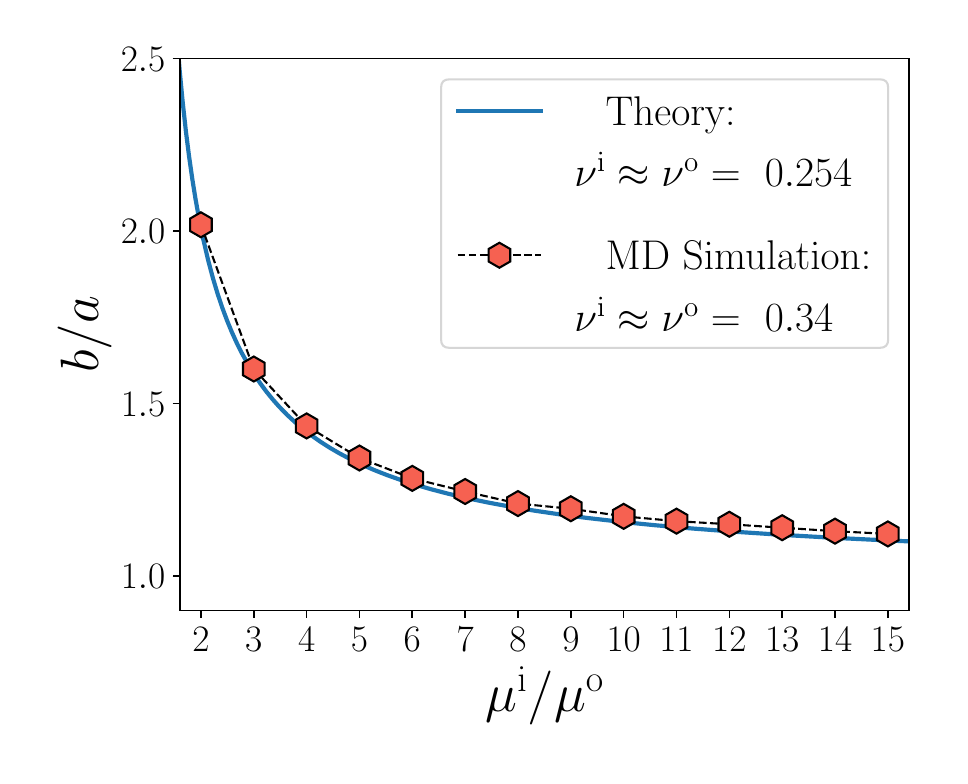}
    }
    \caption{Comparison between the results from continuum elasticity theory and MD simulations. For given materials of $\nu^\mathrm{i}$ $\approx$ $\nu^\mathrm{o}$ $= 0.34$ in two-dimensional MD simulations, we vary the ratio of the elastic moduli between the concealing shell around the enclosed hole and the surrounding elastic solid $\mu^\mathrm{i}/\mu^\mathrm{o}$. We determine the ratio between the radius of the stiff shell $b$ that is necessary to mechanically conceal the presence of the enclosed hole of radius $a$ under uniform compression. For comparison, the corresponding line obtained via Eq.~(\ref{eq:ratio-b-a}) from continuum elasticity theory is included. There, under plane-strain conditions, the effective corresponding three-dimensional Poisson ratios are given by $\nu^\mathrm{i}$ $\approx$ $\nu^\mathrm{o}$ $\approx 0.254$. The results from the two different approaches, continuum theory and discrete microscopic simulations, agree well with each other. (In the MD simulations, we set $a =$ 20~\AA{}, implying a ratio between $a$ and box size $L$ of $a/L = 0.089$.)}
    \label{fig:md_2}
\end{figure}

Quite remarkably, as shown in Fig.~\ref{fig:md_2}, results from atomistic simulations match well the results from analytical continuum elasticity theory over a wide range of ratios $\mu^\mathrm{i}/\mu^\mathrm{o}$ for the considered $\nu^\mathrm{i}$ $\approx$ $\nu^\mathrm{o}$ $=0.34$. This value obtained from the two-dimensional MD simulations converts into $\nu^\mathrm{i}$ $\approx$ $\nu^\mathrm{o}$ $\approx0.254$ for comparison with the theory, as mentioned above. The agreement in Fig.~\ref{fig:md_2} suggests that the picture provided by continuum elasticity theory remains valid even down to atomistic scales for the considered discrete systems of particles interacting solely via LJ pair potentials. 
It confirms our goal of demonstrating a way of mechanical concealment under the given loading conditions that is accessible both on a macroscopic continuum level and down to microscopic, atomistic scales.

\section{Conclusions} 
Summarizing, we considered a cylindrical hole that is surrounded by a stiffer cylindrical shell in a block of elastic material. The material is loaded uniformly under plane-strain conditions. Our scope was to identify for given material parameters a thickness of the shell that mechanically conceals the presence of the hole on the macroscopic level. In other words, the material as a whole, including such masked holes, behaves mechanically in the same way as if the holes were absent. Such a situation allows reduction in weight of components during light-weight construction, maintaining their overall shape and mechanical properties. 

Indeed, a corresponding expression for the thickness of the shell around the hole was identified and derived as a function of the mechanical material parameters and the radius of the concealed hole. For this purpose, we assumed continuous, linearly elastic, homogeneous, isotropic materials under plane-strain conditions. The concentrations of the holes were low enough so that mechanical interactions between the holes can be neglected. 
One step further, we extended this possibility of mechanical concealment from a macroscopic continuum perspective down to discrete atomistic scales using molecular dynamics simulations. A purely atomistic, planar model solid was considered. Overall, consistent results were found. They validate the concept introduced via continuum elasticity theory down to the atomistic level. 
 
As a central step, we thus resolve the concept predicted by continuum elasticity theory on a discrete particulate scale by molecular dynamics simulations. While the continuum elasticity solution renders the problem tractable with only little effort in optimization when adjusting the shell thickness, molecular dynamics simulations motivate broad opportunities for microscopic manipulations that may lead to innovative cloaking techniques.

Future extensions of this work are manifold, important, and obvious. First, we remark that our results within continuum elasticity theory directly carry over to uniform expansion of the system, which, however, may be less abundant in practical situations. An upcoming investigation shall address more general types of loading, beyond uniform. Specifically, our next steps concern uniaxial strains and shear. To lowest order, this extension will allow us to include mutual interactions between shielded holes. Besides plane-strain, also plane-stress and three-dimensional situations shall be considered in future works. Elevated concentrations of holes that make them mutually interact through their induced deformations shall be evaluated. More complex considerations include mechanically anisotropic materials. Altogether, we wish to support the further development of quantitative measures in the context of light-weight materials design.

\begin{acknowledgments}
We thank the European Union (EFRE) and the State Saxony Anhalt for support of this work through project no.\ ZS/2024/02/184030. Moreover, we thank Lukas Fi\-scher, Johannes Menning, and Thomas Wallmersperger for stimulating discussions.
\end{acknowledgments}

\section*{Data Availability Statement}

The generated molecular dynamics simulation data, used to derive the results
presented herein, including the figures, are openly available in the repository Zenodo and can be found at \href{https://doi.org/10.5281/zenodo.18470981}{https://doi.org/10.5281/zenodo.18470981}.

\appendix*

\section{Lennard-Jones ``toy model'' for a strongly cohesive solid with a hole \label{app1}}

Atomistic modeling of holes inside solids is challenging, given that the holes could deform, distort in search of the thermodynamically equilibrated state. Therefore, such geometries are obscure in the literature on MD simulations. 
Numerical investigations of mechanical concealment or cloaking often consider specific materials, such as graphene, using the finite-element approach \citep{buckmann2015mechanical}. 
Furthermore, covalently bonded materials necessitate the usage of many-body interatomic potentials. They can be stabilized with voids in MD simulations, but could invoke anisotropy due to directional bonding and consequently having interatomic potentials with angular dependence.

Choosing specific combinations of materials in atomistic simulations in our specific geometry of a hole, shielded by an elastic shell, embedded in a surrounding elastic solid, limits, for each realization, the comparison between continuum theory and atomistic simulations to a specific ratio of the material parameters $\mu^\mathrm{i}/\mu^\mathrm{o}$. 

\begin{figure}[b]
\centering
 \includegraphics[width=8.5cm]{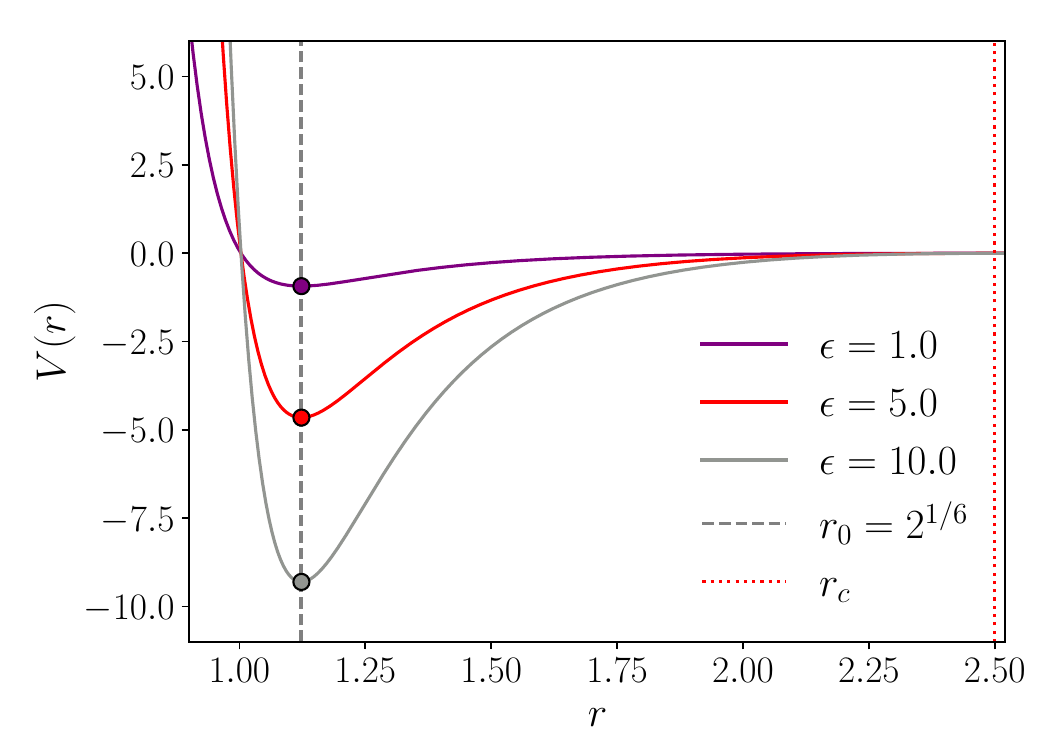}
 \caption{\label{fig:lj} Truncated and shifted Lennard-Jones potential, see Eqs.~(\ref{eq:lj1}) and (\ref{eq:lj2}). The curves illustrate realizations for different values of the parameter $\epsilon$ quantifying the depth of the energy well. Apart from that, the length parameter $\sigma= 1$, equilibrium distance $r_{0} = 2^{1/6}\sigma$ (gray dashed line and bold dots), and cutoff distance $r_c$ (red dotted line) are identical in all cases. In our realization, different stiffness parameters $\epsilon$ are used, depending on whether the interacting atoms are parts of the shell around the hole or of the surrounding elastic solid.}
\end{figure}

To this end, we choose to model a physically traceable system that is simple enough to retain isotropy (potential without angular dependence), yet complex enough to allow mechanical concealment using a shell of variable stiffness around a hole. 

We employ a two-dimensional Lennard-Jones (LJ) solid to mimic a strongly cohesive solid. In Eqs.~(\ref{eq:lj1}) and (\ref{eq:lj2}), we choose $\sigma$ = 1~\AA{} for both types of atoms, part of the shell or the surrounding solid. We modify the energy scale (depth of the energy well), mimicking a ``toy model'' for a strongly cohesive solid with $\epsilon^{\mathrm{o}} = 1$~eV  and $ 2 \epsilon^{\mathrm{o}} \leq \epsilon^{\mathrm{i}} \leq 15 \epsilon^{\mathrm{o}} $, where $\epsilon^\mathrm{i,o}$ quantifies the stiffness between the atoms within the shell ($^\mathrm{i}$) and within the background solid ($^\mathrm{o}$), see Fig.~\ref{fig:lj}. Similar approaches to control the interparticle stiffness using a single parameter in a pairwise potential are found in the literature, albeit in a different context of modeling binary glass formers with a slightly different shape of the pair potential \citep{lerner2022ultrahigh}.

\begin{figure}[b]
\centering
 \includegraphics[width=7cm]{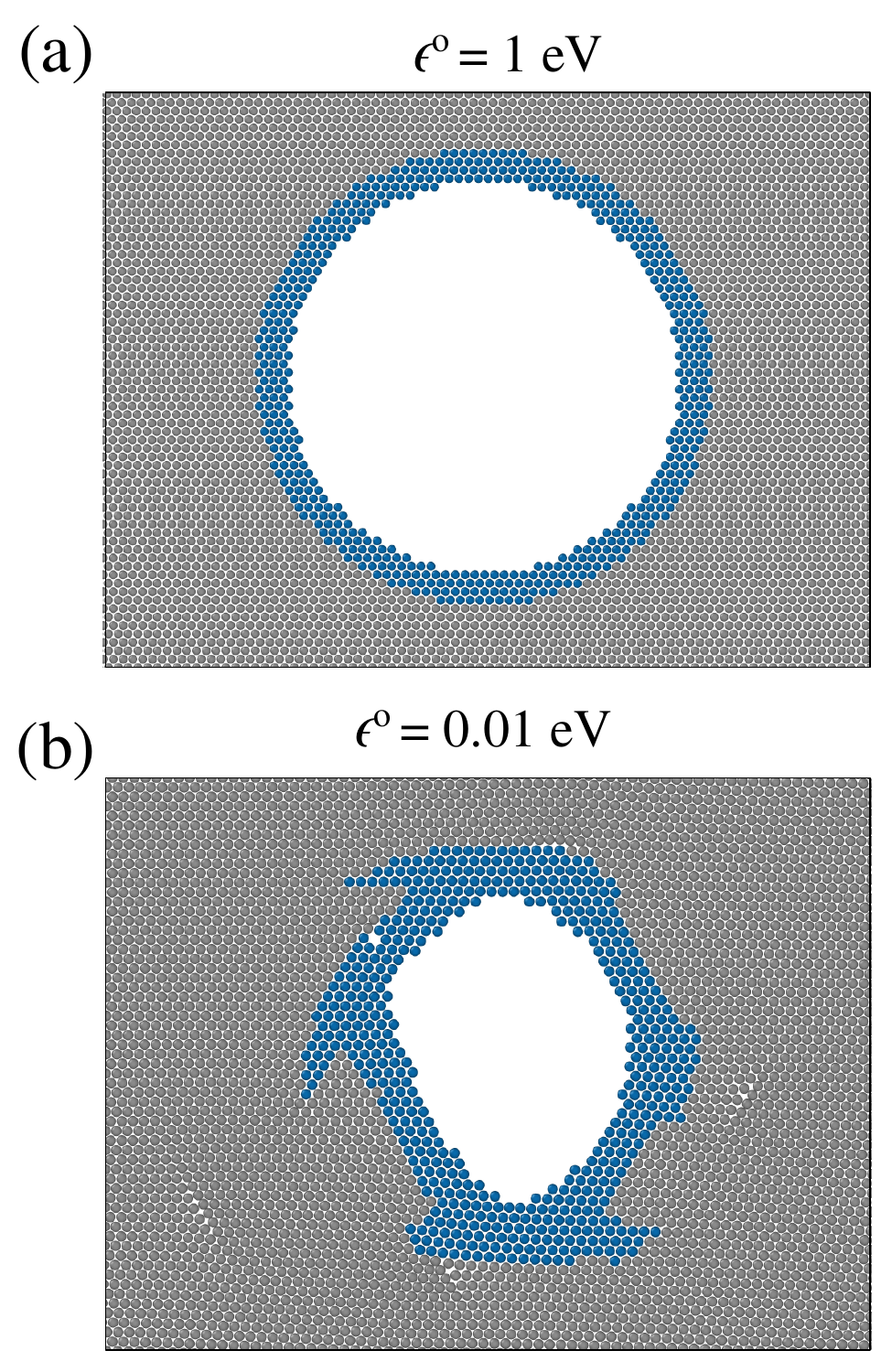}
 \caption{\label{fig:hole} Undistorted and distorted hole under NPT equilibration at $P = 2\times 10^{5}$ bar. We display parts of planar solids with a hole for an LJ energy parameter of the background solid (a) $\epsilon^{\mathrm{o}} = 1$~eV and (b) $\epsilon^{\mathrm{o}} = 0.01$~eV. $\epsilon^{\mathrm{i}}/\epsilon^{\mathrm{o}} = 10$ in both cases.} 

\end{figure}

\begin{figure}[t]
\centering
 \includegraphics[width=7cm]{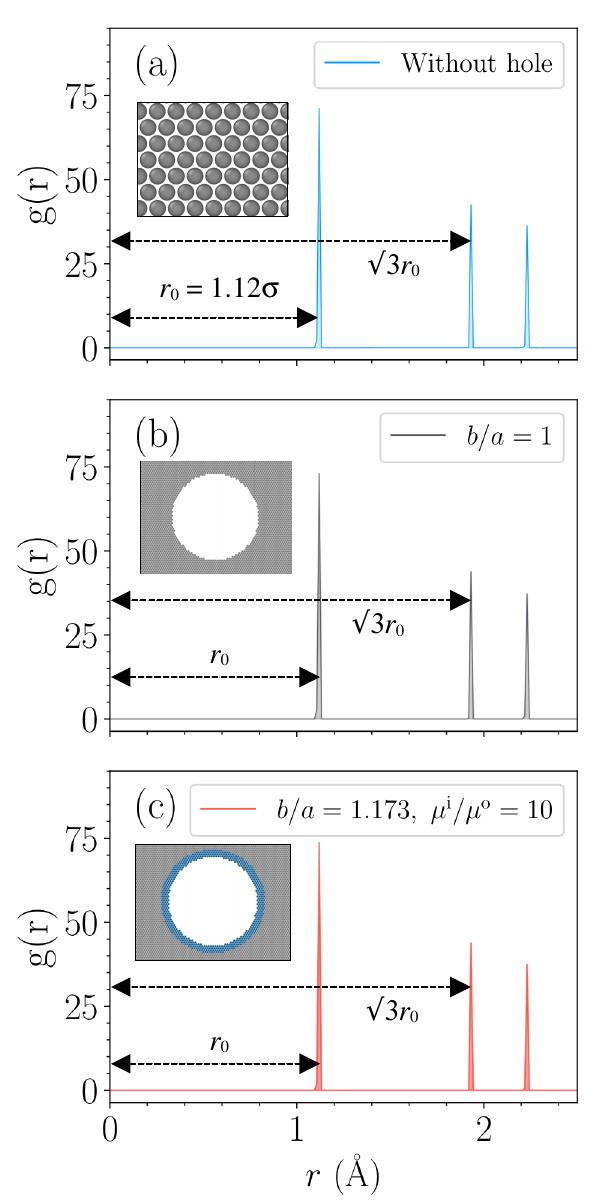}
 \caption{\label{fig:rdf} Radial distribution function g(r) for (a) a pristine planar LJ solid, (b) a solid with an unshielded hole ($b/a = 1$), and (c) a solid with a shielded hole with $b/a = 1.173$ and $\mu^\mathrm{i}/\mu^\mathrm{o} = 10$. Insets show parts of the corresponding atomistic systems. Lattice parameters agree with the corresponding minima of the truncated and shifted LJ energies ($r_{0} = 1.12 \sigma$). The second-neighbor distances are consistent with hexagonal symmetry.}
\end{figure}

\begin{table} [!htb]
\caption{\label{tab:table1} Positions and corresponding heights of the peaks of the radial distribution function g(r) for a pristine planar LJ solid, a solid with an unshielded hole, and a solid with a shielded hole, rounded to decimals as listed.\\}
\begin{ruledtabular}
\begin{tabular}{c|ccc|ccc}
 &\multicolumn{3}{c}{g(r) peak positions (\AA{})} &\multicolumn{3}{c}{g(r) peak heights}\\

System & Peak 1 & Peak 2 & Peak 3  & Peak 1 & Peak 2 & Peak 3\\
\hline
Pristine & 1.119 & 1.931 & 2.231 & 71.25 & 42.66 & 36.38 \\
\hline
Unshielded hole & 1.119 & 1.931 & 2.231 & 73.0 & 43.8 & 37.3  \\
($b/a$ = 1) &  &  &  &  & \\
\hline
Shielded hole & 1.119 & 1.931 & 2.231 & 73.8 & 43.8 & 37.8  \\
($b/a$ = 1.173) &  &  &  &  & \\
\end{tabular}
\end{ruledtabular}
\end{table}

Indeed, stability of a LJ solid in two dimensions with a void is unattainable using the standard cohesive energy values of rare-gas solids ($\epsilon \sim 10^{-2}$~eV for Ar). Therefore, we use deeper energy wells, which results in very high elastic moduli of the order of TPa. We remark, that monolayer graphene, having a honeycomb hexagonal structure, possesses similarly high elastic moduli of the order of TPa. Our two-dimensional LJ model of a strongly cohesive solid is even stiffer. In our case, we study a hexagonal, six-neighbor structure, as opposed to the three-neighbor structure in monolayer graphene.

Overall, we thus remark that the parameters for our LJ solid were adjusted to maintain a simple, basic description as a proof of concept on the atomic scale. Although, the considered parameters do not reflect a realistic solid at these elevated values for the elastic moduli, we note that the LJ energy parameters $\epsilon$ for metals \citep{jacobson2022revisting, jacobson2022corrigendum, filippova2015calculation} and even for two-dimensional materials \citep{momeni2020multiscale} often correspond to such high values ranging mostly within $0.1$~eV $< \epsilon < 1.5$~eV.

As an example, we include a relative comparison between $\epsilon^{\mathrm{o}} = 0.01$ and 1~eV in sustaining a shielded hole under NPT equilibration at $P = 2\times 10^{5}$ bar in Fig.~\ref{fig:hole}. Reducing $\epsilon^{\mathrm{o}}$ to 0.1~eV can sustain the hole with slightly lower bulk modulus. Yet, we chose to use $\epsilon^{\mathrm{o}} = 1$ eV and $\sigma$ = 1~\AA{} so that the problem can be conveniently expressed in reduced units. Similarly, larger values of $\epsilon$ were used to simulate two-dimensional LJ solids with periodically arranged square holes in the context of stress analysis in inhomogeneous materials \citep{rigelesaiyin2018asymmetry}.

Consistently with the chosen value of $\sigma$, all atomistic systems simulated in this work, irrespective of whether they contain holes (shielded and unshielded) or not, possess a constant lattice parameter $r_{0} = 2^{1/6}\sigma$. This can be inferred from the first peak position of the radial distribution function g(r) computed for different systems with and without holes, see Fig.~\ref{fig:rdf} and Table~\ref{tab:table1}. Additionally, Table~\ref{tab:table1} presents the positions and corresponding heights of the second and third peaks of g(r). This confirmation is crucial for implementing the mechanical cloaking, as we rely on the relation $\mu \sim \epsilon/r_{0}^{2}$ in fixing the shear moduli for the background ($\mu^{\mathrm{o}}$) and the shielded region ($\mu^{\mathrm{i}}$) within the solid. Therefore, any variation in $r_{0}$ affects the cloaking thickness. Moreover, Fig.~\ref{fig:rdf} as well as Table~\ref{tab:table1} show that the hexagonal symmetry is maintained (see the second-neighbor peak at $\sqrt{3}r_0$) up to the interaction range of the interatomic potential (2.5$\sigma$).

\providecommand{\noopsort}[1]{}\providecommand{\singleletter}[1]{#1}%
%


\begin{thebibliography}{79}%
\makeatletter
\providecommand \@ifxundefined [1]{%
 \@ifx{#1\undefined}
}%
\providecommand \@ifnum [1]{%
 \ifnum #1\expandafter \@firstoftwo
 \else \expandafter \@secondoftwo
 \fi
}%
\providecommand \@ifx [1]{%
 \ifx #1\expandafter \@firstoftwo
 \else \expandafter \@secondoftwo
 \fi
}%
\providecommand \natexlab [1]{#1}%
\providecommand \enquote  [1]{``#1''}%
\providecommand \bibnamefont  [1]{#1}%
\providecommand \bibfnamefont [1]{#1}%
\providecommand \citenamefont [1]{#1}%
\providecommand \href@noop [0]{\@secondoftwo}%
\providecommand \href [0]{\begingroup \@sanitize@url \@href}%
\providecommand \@href[1]{\@@startlink{#1}\@@href}%
\providecommand \@@href[1]{\endgroup#1\@@endlink}%
\providecommand \@sanitize@url [0]{\catcode `\\12\catcode `\$12\catcode `\&12\catcode `\#12\catcode `\^12\catcode `\_12\catcode `\%12\relax}%
\providecommand \@@startlink[1]{}%
\providecommand \@@endlink[0]{}%
\providecommand \url  [0]{\begingroup\@sanitize@url \@url }%
\providecommand \@url [1]{\endgroup\@href {#1}{\urlprefix }}%
\providecommand \urlprefix  [0]{URL }%
\providecommand \Eprint [0]{\href }%
\providecommand \doibase [0]{http://dx.doi.org/}%
\providecommand \selectlanguage [0]{\@gobble}%
\providecommand \bibinfo  [0]{\@secondoftwo}%
\providecommand \bibfield  [0]{\@secondoftwo}%
\providecommand \translation [1]{[#1]}%
\providecommand \BibitemOpen [0]{}%
\providecommand \bibitemStop [0]{}%
\providecommand \bibitemNoStop [0]{.\EOS\space}%
\providecommand \EOS [0]{\spacefactor3000\relax}%
\providecommand \BibitemShut  [1]{\csname bibitem#1\endcsname}%
\let\auto@bib@innerbib\@empty
\bibitem [{\citenamefont {K{\"o}nig}, \citenamefont {Mathieu},\ and\ \citenamefont {Vielhaber}(2024)}]{konig2024resource}%
  \BibitemOpen
  \bibfield  {author} {\bibinfo {author} {\bibfnamefont {K.}~\bibnamefont {K{\"o}nig}}, \bibinfo {author} {\bibfnamefont {J.}~\bibnamefont {Mathieu}}, \ and\ \bibinfo {author} {\bibfnamefont {M.}~\bibnamefont {Vielhaber}},\ }\bibfield  {title} {\enquote {\bibinfo {title} {Resource conservation by means of lightweight design and design for circularity—a concept for decision making in the early phase of product development},}\ }\href@noop {} {\bibfield  {journal} {\bibinfo  {journal} {Res. Conserv. Recycl.}\ }\textbf {\bibinfo {volume} {201}},\ \bibinfo {pages} {107331} (\bibinfo {year} {2024})}\BibitemShut {NoStop}%
\bibitem [{\citenamefont {Lai}\ \emph {et~al.}(2015)\citenamefont {Lai}, \citenamefont {Price}, \citenamefont {Modla}, \citenamefont {Thompson}, \citenamefont {Caplan}, \citenamefont {Kirn-Safran},\ and\ \citenamefont {Wang}}]{lai2015dependences}%
  \BibitemOpen
  \bibfield  {author} {\bibinfo {author} {\bibfnamefont {X.}~\bibnamefont {Lai}}, \bibinfo {author} {\bibfnamefont {C.}~\bibnamefont {Price}}, \bibinfo {author} {\bibfnamefont {S.}~\bibnamefont {Modla}}, \bibinfo {author} {\bibfnamefont {W.~R.}\ \bibnamefont {Thompson}}, \bibinfo {author} {\bibfnamefont {J.}~\bibnamefont {Caplan}}, \bibinfo {author} {\bibfnamefont {C.~B.}\ \bibnamefont {Kirn-Safran}}, \ and\ \bibinfo {author} {\bibfnamefont {L.}~\bibnamefont {Wang}},\ }\bibfield  {title} {\enquote {\bibinfo {title} {The dependences of osteocyte network on bone compartment, age, and disease},}\ }\href@noop {} {\bibfield  {journal} {\bibinfo  {journal} {Bone Res.}\ }\textbf {\bibinfo {volume} {3}},\ \bibinfo {pages} {1} (\bibinfo {year} {2015})}\BibitemShut {NoStop}%
\bibitem [{\citenamefont {Yu}\ \emph {et~al.}(2020)\citenamefont {Yu}, \citenamefont {Pacureanu}, \citenamefont {Olivier}, \citenamefont {Cloetens},\ and\ \citenamefont {Peyrin}}]{yu2020assessment}%
  \BibitemOpen
  \bibfield  {author} {\bibinfo {author} {\bibfnamefont {B.}~\bibnamefont {Yu}}, \bibinfo {author} {\bibfnamefont {A.}~\bibnamefont {Pacureanu}}, \bibinfo {author} {\bibfnamefont {C.}~\bibnamefont {Olivier}}, \bibinfo {author} {\bibfnamefont {P.}~\bibnamefont {Cloetens}}, \ and\ \bibinfo {author} {\bibfnamefont {F.}~\bibnamefont {Peyrin}},\ }\bibfield  {title} {\enquote {\bibinfo {title} {Assessment of the human bone lacuno-canalicular network at the nanoscale and impact of spatial resolution},}\ }\href@noop {} {\bibfield  {journal} {\bibinfo  {journal} {Sci. Rep.}\ }\textbf {\bibinfo {volume} {10}},\ \bibinfo {pages} {4567} (\bibinfo {year} {2020})}\BibitemShut {NoStop}%
\bibitem [{\citenamefont {Koffler}\ and\ \citenamefont {Rohde-Brandenburger}(2010)}]{koffler2010calculation}%
  \BibitemOpen
  \bibfield  {author} {\bibinfo {author} {\bibfnamefont {C.}~\bibnamefont {Koffler}}\ and\ \bibinfo {author} {\bibfnamefont {K.}~\bibnamefont {Rohde-Brandenburger}},\ }\bibfield  {title} {\enquote {\bibinfo {title} {On the calculation of fuel savings through lightweight design in automotive life cycle assessments},}\ }\href@noop {} {\bibfield  {journal} {\bibinfo  {journal} {Int. J. Life Cycle Assess.}\ }\textbf {\bibinfo {volume} {15}},\ \bibinfo {pages} {128} (\bibinfo {year} {2010})}\BibitemShut {NoStop}%
\bibitem [{\citenamefont {Timmis}\ \emph {et~al.}(2015)\citenamefont {Timmis}, \citenamefont {Hodzic}, \citenamefont {Koh}, \citenamefont {Bonner}, \citenamefont {Soutis}, \citenamefont {Sch{\"a}fer},\ and\ \citenamefont {Dray}}]{timmis2015environmental}%
  \BibitemOpen
  \bibfield  {author} {\bibinfo {author} {\bibfnamefont {A.~J.}\ \bibnamefont {Timmis}}, \bibinfo {author} {\bibfnamefont {A.}~\bibnamefont {Hodzic}}, \bibinfo {author} {\bibfnamefont {L.}~\bibnamefont {Koh}}, \bibinfo {author} {\bibfnamefont {M.}~\bibnamefont {Bonner}}, \bibinfo {author} {\bibfnamefont {C.}~\bibnamefont {Soutis}}, \bibinfo {author} {\bibfnamefont {A.~W.}\ \bibnamefont {Sch{\"a}fer}}, \ and\ \bibinfo {author} {\bibfnamefont {L.}~\bibnamefont {Dray}},\ }\bibfield  {title} {\enquote {\bibinfo {title} {Environmental impact assessment of aviation emission reduction through the implementation of composite materials},}\ }\href@noop {} {\bibfield  {journal} {\bibinfo  {journal} {Int. J. Life Cycle Assess.}\ }\textbf {\bibinfo {volume} {20}},\ \bibinfo {pages} {233} (\bibinfo {year} {2015})}\BibitemShut {NoStop}%
\bibitem [{\citenamefont {Wang}\ and\ \citenamefont {Salhab}(2009)}]{wang2009structural}%
  \BibitemOpen
  \bibfield  {author} {\bibinfo {author} {\bibfnamefont {Y.}~\bibnamefont {Wang}}\ and\ \bibinfo {author} {\bibfnamefont {B.}~\bibnamefont {Salhab}},\ }\bibfield  {title} {\enquote {\bibinfo {title} {Structural behaviour and design of lightweight structural panels using perforated cold-formed thin-walled sections under compression},}\ }\href@noop {} {\bibfield  {journal} {\bibinfo  {journal} {Int. J. Steel Struct.}\ }\textbf {\bibinfo {volume} {9}},\ \bibinfo {pages} {57} (\bibinfo {year} {2009})}\BibitemShut {NoStop}%
\bibitem [{\citenamefont {Zhou}\ \emph {et~al.}(2022)\citenamefont {Zhou}, \citenamefont {Zhuang}, \citenamefont {Hu}, \citenamefont {Hu}, \citenamefont {Huang}, \citenamefont {Li}, \citenamefont {Guo},\ and\ \citenamefont {Zhu}}]{zhou2022perforated}%
  \BibitemOpen
  \bibfield  {author} {\bibinfo {author} {\bibfnamefont {Y.}~\bibnamefont {Zhou}}, \bibinfo {author} {\bibfnamefont {L.}~\bibnamefont {Zhuang}}, \bibinfo {author} {\bibfnamefont {Z.}~\bibnamefont {Hu}}, \bibinfo {author} {\bibfnamefont {B.}~\bibnamefont {Hu}}, \bibinfo {author} {\bibfnamefont {X.}~\bibnamefont {Huang}}, \bibinfo {author} {\bibfnamefont {S.}~\bibnamefont {Li}}, \bibinfo {author} {\bibfnamefont {M.}~\bibnamefont {Guo}}, \ and\ \bibinfo {author} {\bibfnamefont {Z.}~\bibnamefont {Zhu}},\ }\bibfield  {title} {\enquote {\bibinfo {title} {Perforated steel block of realizing large ductility under compression: Parametric study and stress--strain modeling},}\ }\href@noop {} {\bibfield  {journal} {\bibinfo  {journal} {Rev. Adv. Mater. Sci.}\ }\textbf {\bibinfo {volume} {61}},\ \bibinfo {pages} {221} (\bibinfo {year} {2022})}\BibitemShut {NoStop}%
\bibitem [{\citenamefont {Bruggi}\ \emph {et~al.}(2022)\citenamefont {Bruggi}, \citenamefont {Ismail}, \citenamefont {L{\'o}g{\'o}},\ and\ \citenamefont {Paoletti}}]{bruggi2022lightweight}%
  \BibitemOpen
  \bibfield  {author} {\bibinfo {author} {\bibfnamefont {M.}~\bibnamefont {Bruggi}}, \bibinfo {author} {\bibfnamefont {H.}~\bibnamefont {Ismail}}, \bibinfo {author} {\bibfnamefont {J.}~\bibnamefont {L{\'o}g{\'o}}}, \ and\ \bibinfo {author} {\bibfnamefont {I.}~\bibnamefont {Paoletti}},\ }\bibfield  {title} {\enquote {\bibinfo {title} {Lightweight design with displacement constraints using graded porous microstructures},}\ }\href@noop {} {\bibfield  {journal} {\bibinfo  {journal} {Comput. Struct.}\ }\textbf {\bibinfo {volume} {272}},\ \bibinfo {pages} {106873} (\bibinfo {year} {2022})}\BibitemShut {NoStop}%
\bibitem [{\citenamefont {Lu}, \citenamefont {Xu},\ and\ \citenamefont {Zhang}(2017)}]{lu2017stress}%
  \BibitemOpen
  \bibfield  {author} {\bibinfo {author} {\bibfnamefont {A.-Z.}\ \bibnamefont {Lu}}, \bibinfo {author} {\bibfnamefont {Z.}~\bibnamefont {Xu}}, \ and\ \bibinfo {author} {\bibfnamefont {N.}~\bibnamefont {Zhang}},\ }\bibfield  {title} {\enquote {\bibinfo {title} {Stress analytical solution for an infinite plane containing two holes},}\ }\href@noop {} {\bibfield  {journal} {\bibinfo  {journal} {Int. J. Mech. Sci.}\ }\textbf {\bibinfo {volume} {128}},\ \bibinfo {pages} {224} (\bibinfo {year} {2017})}\BibitemShut {NoStop}%
\bibitem [{\citenamefont {Liu}(2006)}]{liu2006new}%
  \BibitemOpen
  \bibfield  {author} {\bibinfo {author} {\bibfnamefont {Y.}~\bibnamefont {Liu}},\ }\bibfield  {title} {\enquote {\bibinfo {title} {A new fast multipole boundary element method for solving large-scale two-dimensional elastostatic problems},}\ }\href@noop {} {\bibfield  {journal} {\bibinfo  {journal} {Int. J. Numer. Meth. Engng.}\ }\textbf {\bibinfo {volume} {65}},\ \bibinfo {pages} {863} (\bibinfo {year} {2006})}\BibitemShut {NoStop}%
\bibitem [{\citenamefont {Kang}, \citenamefont {Wang},\ and\ \citenamefont {Tan}(2018)}]{kang2018cavitation}%
  \BibitemOpen
  \bibfield  {author} {\bibinfo {author} {\bibfnamefont {J.}~\bibnamefont {Kang}}, \bibinfo {author} {\bibfnamefont {C.}~\bibnamefont {Wang}}, \ and\ \bibinfo {author} {\bibfnamefont {H.}~\bibnamefont {Tan}},\ }\bibfield  {title} {\enquote {\bibinfo {title} {Cavitation in inhomogeneous soft solids},}\ }\href@noop {} {\bibfield  {journal} {\bibinfo  {journal} {Soft Matter}\ }\textbf {\bibinfo {volume} {14}},\ \bibinfo {pages} {7979} (\bibinfo {year} {2018})}\BibitemShut {NoStop}%
\bibitem [{\citenamefont {Kang}\ and\ \citenamefont {Tang}(2021)}]{kang2021dynamic}%
  \BibitemOpen
  \bibfield  {author} {\bibinfo {author} {\bibfnamefont {J.}~\bibnamefont {Kang}}\ and\ \bibinfo {author} {\bibfnamefont {Y.}~\bibnamefont {Tang}},\ }\bibfield  {title} {\enquote {\bibinfo {title} {Dynamic cavitation in soft solids under monotonically increasing pressure},}\ }\href@noop {} {\bibfield  {journal} {\bibinfo  {journal} {Int. J. Mech. Sci.}\ }\textbf {\bibinfo {volume} {209}},\ \bibinfo {pages} {106730} (\bibinfo {year} {2021})}\BibitemShut {NoStop}%
\bibitem [{\citenamefont {Cheng}\ \emph {et~al.}(2023)\citenamefont {Cheng}, \citenamefont {Liu}, \citenamefont {Zhang}, \citenamefont {Tang}, \citenamefont {Liu}, \citenamefont {Tang}, \citenamefont {Du}, \citenamefont {Cui},\ and\ \citenamefont {Guo}}]{cheng2023compatible}%
  \BibitemOpen
  \bibfield  {author} {\bibinfo {author} {\bibfnamefont {X.}~\bibnamefont {Cheng}}, \bibinfo {author} {\bibfnamefont {C.}~\bibnamefont {Liu}}, \bibinfo {author} {\bibfnamefont {W.}~\bibnamefont {Zhang}}, \bibinfo {author} {\bibfnamefont {Z.}~\bibnamefont {Tang}}, \bibinfo {author} {\bibfnamefont {Y.}~\bibnamefont {Liu}}, \bibinfo {author} {\bibfnamefont {S.}~\bibnamefont {Tang}}, \bibinfo {author} {\bibfnamefont {Z.}~\bibnamefont {Du}}, \bibinfo {author} {\bibfnamefont {T.}~\bibnamefont {Cui}}, \ and\ \bibinfo {author} {\bibfnamefont {X.}~\bibnamefont {Guo}},\ }\bibfield  {title} {\enquote {\bibinfo {title} {A compatible boundary condition-based topology optimization paradigm for static mechanical cloak design},}\ }\href@noop {} {\bibfield  {journal} {\bibinfo  {journal} {Extrem. Mech. Lett.}\ }\textbf {\bibinfo {volume} {65}},\ \bibinfo {pages} {102100} (\bibinfo {year} {2023})}\BibitemShut {NoStop}%
\bibitem [{\citenamefont {B{\"u}ckmann}\ \emph {et~al.}(2014)\citenamefont {B{\"u}ckmann}, \citenamefont {Thiel}, \citenamefont {Kadic}, \citenamefont {Schittny},\ and\ \citenamefont {Wegener}}]{buckmann2014elasto}%
  \BibitemOpen
  \bibfield  {author} {\bibinfo {author} {\bibfnamefont {T.}~\bibnamefont {B{\"u}ckmann}}, \bibinfo {author} {\bibfnamefont {M.}~\bibnamefont {Thiel}}, \bibinfo {author} {\bibfnamefont {M.}~\bibnamefont {Kadic}}, \bibinfo {author} {\bibfnamefont {R.}~\bibnamefont {Schittny}}, \ and\ \bibinfo {author} {\bibfnamefont {M.}~\bibnamefont {Wegener}},\ }\bibfield  {title} {\enquote {\bibinfo {title} {An elasto-mechanical unfeelability cloak made of pentamode metamaterials},}\ }\href@noop {} {\bibfield  {journal} {\bibinfo  {journal} {Nat. Commun.}\ }\textbf {\bibinfo {volume} {5}},\ \bibinfo {pages} {4130} (\bibinfo {year} {2014})}\BibitemShut {NoStop}%
\bibitem [{\citenamefont {Diatta}\ and\ \citenamefont {Guenneau}(2014)}]{diatta2014controlling}%
  \BibitemOpen
  \bibfield  {author} {\bibinfo {author} {\bibfnamefont {A.}~\bibnamefont {Diatta}}\ and\ \bibinfo {author} {\bibfnamefont {S.}~\bibnamefont {Guenneau}},\ }\bibfield  {title} {\enquote {\bibinfo {title} {Controlling solid elastic waves with spherical cloaks},}\ }\href@noop {} {\bibfield  {journal} {\bibinfo  {journal} {Appl. Phys. Lett.}\ }\textbf {\bibinfo {volume} {105}},\ \bibinfo {pages} {021901} (\bibinfo {year} {2014})}\BibitemShut {NoStop}%
\bibitem [{\citenamefont {Achaoui}\ \emph {et~al.}(2020)\citenamefont {Achaoui}, \citenamefont {Diatta}, \citenamefont {Kadic},\ and\ \citenamefont {Guenneau}}]{achaoui2020cloaking}%
  \BibitemOpen
  \bibfield  {author} {\bibinfo {author} {\bibfnamefont {Y.}~\bibnamefont {Achaoui}}, \bibinfo {author} {\bibfnamefont {A.}~\bibnamefont {Diatta}}, \bibinfo {author} {\bibfnamefont {M.}~\bibnamefont {Kadic}}, \ and\ \bibinfo {author} {\bibfnamefont {S.}~\bibnamefont {Guenneau}},\ }\bibfield  {title} {\enquote {\bibinfo {title} {Cloaking in-plane elastic waves with swiss rolls},}\ }\href@noop {} {\bibfield  {journal} {\bibinfo  {journal} {Materials}\ }\textbf {\bibinfo {volume} {13}},\ \bibinfo {pages} {449} (\bibinfo {year} {2020})}\BibitemShut {NoStop}%
\bibitem [{\citenamefont {Stenger}, \citenamefont {Wilhelm},\ and\ \citenamefont {Wegener}(2012)}]{stenger2012experiments}%
  \BibitemOpen
  \bibfield  {author} {\bibinfo {author} {\bibfnamefont {N.}~\bibnamefont {Stenger}}, \bibinfo {author} {\bibfnamefont {M.}~\bibnamefont {Wilhelm}}, \ and\ \bibinfo {author} {\bibfnamefont {M.}~\bibnamefont {Wegener}},\ }\bibfield  {title} {\enquote {\bibinfo {title} {Experiments on elastic cloaking in thin plates},}\ }\href@noop {} {\bibfield  {journal} {\bibinfo  {journal} {Phys. Rev. Lett.}\ }\textbf {\bibinfo {volume} {108}},\ \bibinfo {pages} {014301} (\bibinfo {year} {2012})}\BibitemShut {NoStop}%
\bibitem [{\citenamefont {Cheng}\ \emph {et~al.}(2025)\citenamefont {Cheng}, \citenamefont {Du}, \citenamefont {Meng}, \citenamefont {Liu}, \citenamefont {Zhang},\ and\ \citenamefont {Guo}}]{cheng2025thermomechanical}%
  \BibitemOpen
  \bibfield  {author} {\bibinfo {author} {\bibfnamefont {X.}~\bibnamefont {Cheng}}, \bibinfo {author} {\bibfnamefont {Z.}~\bibnamefont {Du}}, \bibinfo {author} {\bibfnamefont {W.}~\bibnamefont {Meng}}, \bibinfo {author} {\bibfnamefont {C.}~\bibnamefont {Liu}}, \bibinfo {author} {\bibfnamefont {W.}~\bibnamefont {Zhang}}, \ and\ \bibinfo {author} {\bibfnamefont {X.}~\bibnamefont {Guo}},\ }\bibfield  {title} {\enquote {\bibinfo {title} {Thermomechanical cloaking via compatibility-driven topology optimization},}\ }\href@noop {} {\bibfield  {journal} {\bibinfo  {journal} {Int. J. Mech. Sci.}\ }\textbf {\bibinfo {volume} {300}},\ \bibinfo {pages} {110400} (\bibinfo {year} {2025})}\BibitemShut {NoStop}%
\bibitem [{\citenamefont {Meirbekova}\ and\ \citenamefont {Brun}(2020)}]{meirbekova2020control}%
  \BibitemOpen
  \bibfield  {author} {\bibinfo {author} {\bibfnamefont {B.}~\bibnamefont {Meirbekova}}\ and\ \bibinfo {author} {\bibfnamefont {M.}~\bibnamefont {Brun}},\ }\bibfield  {title} {\enquote {\bibinfo {title} {Control of elastic shear waves by periodic geometric transformation: cloaking, high reflectivity and anomalous resonances},}\ }\href@noop {} {\bibfield  {journal} {\bibinfo  {journal} {J. Mech. Phys. Solids}\ }\textbf {\bibinfo {volume} {137}},\ \bibinfo {pages} {103816} (\bibinfo {year} {2020})}\BibitemShut {NoStop}%
\bibitem [{\citenamefont {B{\"u}ckmann}\ \emph {et~al.}(2015)\citenamefont {B{\"u}ckmann}, \citenamefont {Kadic}, \citenamefont {Schittny},\ and\ \citenamefont {Wegener}}]{buckmann2015mechanical}%
  \BibitemOpen
  \bibfield  {author} {\bibinfo {author} {\bibfnamefont {T.}~\bibnamefont {B{\"u}ckmann}}, \bibinfo {author} {\bibfnamefont {M.}~\bibnamefont {Kadic}}, \bibinfo {author} {\bibfnamefont {R.}~\bibnamefont {Schittny}}, \ and\ \bibinfo {author} {\bibfnamefont {M.}~\bibnamefont {Wegener}},\ }\bibfield  {title} {\enquote {\bibinfo {title} {Mechanical cloak design by direct lattice transformation},}\ }\href@noop {} {\bibfield  {journal} {\bibinfo  {journal} {Proc. Natl. Acad. Sci.}\ }\textbf {\bibinfo {volume} {112}},\ \bibinfo {pages} {4930} (\bibinfo {year} {2015})}\BibitemShut {NoStop}%
\bibitem [{\citenamefont {Wang}\ \emph {et~al.}(2022)\citenamefont {Wang}, \citenamefont {Boddapati}, \citenamefont {Liu}, \citenamefont {Zhu}, \citenamefont {Daraio},\ and\ \citenamefont {Chen}}]{wang2022mechanical}%
  \BibitemOpen
  \bibfield  {author} {\bibinfo {author} {\bibfnamefont {L.}~\bibnamefont {Wang}}, \bibinfo {author} {\bibfnamefont {J.}~\bibnamefont {Boddapati}}, \bibinfo {author} {\bibfnamefont {K.}~\bibnamefont {Liu}}, \bibinfo {author} {\bibfnamefont {P.}~\bibnamefont {Zhu}}, \bibinfo {author} {\bibfnamefont {C.}~\bibnamefont {Daraio}}, \ and\ \bibinfo {author} {\bibfnamefont {W.}~\bibnamefont {Chen}},\ }\bibfield  {title} {\enquote {\bibinfo {title} {Mechanical cloak via data-driven aperiodic metamaterial design},}\ }\href@noop {} {\bibfield  {journal} {\bibinfo  {journal} {Proc. Natl. Acad. Sci.}\ }\textbf {\bibinfo {volume} {119}},\ \bibinfo {pages} {e2122185119} (\bibinfo {year} {2022})}\BibitemShut {NoStop}%
\bibitem [{\citenamefont {Sanders}, \citenamefont {Aguil{\'o}},\ and\ \citenamefont {Paulino}(2021)}]{sanders2021optimized}%
  \BibitemOpen
  \bibfield  {author} {\bibinfo {author} {\bibfnamefont {E.}~\bibnamefont {Sanders}}, \bibinfo {author} {\bibfnamefont {M.}~\bibnamefont {Aguil{\'o}}}, \ and\ \bibinfo {author} {\bibfnamefont {G.}~\bibnamefont {Paulino}},\ }\bibfield  {title} {\enquote {\bibinfo {title} {Optimized lattice-based metamaterials for elastostatic cloaking},}\ }\href@noop {} {\bibfield  {journal} {\bibinfo  {journal} {Proc. R. Soc. A}\ }\textbf {\bibinfo {volume} {477}},\ \bibinfo {pages} {20210418} (\bibinfo {year} {2021})}\BibitemShut {NoStop}%
\bibitem [{\citenamefont {Liu}\ \emph {et~al.}(2025)\citenamefont {Liu}, \citenamefont {Xu}, \citenamefont {Qu}, \citenamefont {Xiao}, \citenamefont {Tao},\ and\ \citenamefont {Gao}}]{liu2025computationally}%
  \BibitemOpen
  \bibfield  {author} {\bibinfo {author} {\bibfnamefont {X.}~\bibnamefont {Liu}}, \bibinfo {author} {\bibfnamefont {J.}~\bibnamefont {Xu}}, \bibinfo {author} {\bibfnamefont {M.}~\bibnamefont {Qu}}, \bibinfo {author} {\bibfnamefont {M.}~\bibnamefont {Xiao}}, \bibinfo {author} {\bibfnamefont {G.}~\bibnamefont {Tao}}, \ and\ \bibinfo {author} {\bibfnamefont {L.}~\bibnamefont {Gao}},\ }\bibfield  {title} {\enquote {\bibinfo {title} {Computationally generative mechanical cloaks},}\ }\href@noop {} {\bibfield  {journal} {\bibinfo  {journal} {Cell Rep. Phys. Sci.}\ }\textbf {\bibinfo {volume} {6}},\ \bibinfo {pages} {102971} (\bibinfo {year} {2025})}\BibitemShut {NoStop}%
\bibitem [{\citenamefont {Craster}\ \emph {et~al.}(2023)\citenamefont {Craster}, \citenamefont {Guenneau}, \citenamefont {Kadic},\ and\ \citenamefont {Wegener}}]{craster2023mechanical}%
  \BibitemOpen
  \bibfield  {author} {\bibinfo {author} {\bibfnamefont {R.}~\bibnamefont {Craster}}, \bibinfo {author} {\bibfnamefont {S.}~\bibnamefont {Guenneau}}, \bibinfo {author} {\bibfnamefont {M.}~\bibnamefont {Kadic}}, \ and\ \bibinfo {author} {\bibfnamefont {M.}~\bibnamefont {Wegener}},\ }\bibfield  {title} {\enquote {\bibinfo {title} {Mechanical metamaterials},}\ }\href@noop {} {\bibfield  {journal} {\bibinfo  {journal} {Rep. Prog. Phys.}\ }\textbf {\bibinfo {volume} {86}},\ \bibinfo {pages} {094501} (\bibinfo {year} {2023})}\BibitemShut {NoStop}%
\bibitem [{\citenamefont {Nassar}, \citenamefont {Chen},\ and\ \citenamefont {Huang}(2018)}]{nassar2018degenerate}%
  \BibitemOpen
  \bibfield  {author} {\bibinfo {author} {\bibfnamefont {H.}~\bibnamefont {Nassar}}, \bibinfo {author} {\bibfnamefont {Y.}~\bibnamefont {Chen}}, \ and\ \bibinfo {author} {\bibfnamefont {G.}~\bibnamefont {Huang}},\ }\bibfield  {title} {\enquote {\bibinfo {title} {A degenerate polar lattice for cloaking in full two-dimensional elastodynamics and statics},}\ }\href@noop {} {\bibfield  {journal} {\bibinfo  {journal} {Proc. R. Soc. A}\ }\textbf {\bibinfo {volume} {474}},\ \bibinfo {pages} {20180523} (\bibinfo {year} {2018})}\BibitemShut {NoStop}%
\bibitem [{\citenamefont {Zhang}\ \emph {et~al.}(2024)\citenamefont {Zhang}, \citenamefont {Tan}, \citenamefont {Shen}, \citenamefont {Yang}, \citenamefont {Peng}, \citenamefont {Liu},\ and\ \citenamefont {Du}}]{zhang2024realizable}%
  \BibitemOpen
  \bibfield  {author} {\bibinfo {author} {\bibfnamefont {Y.}~\bibnamefont {Zhang}}, \bibinfo {author} {\bibfnamefont {J.}~\bibnamefont {Tan}}, \bibinfo {author} {\bibfnamefont {Y.}~\bibnamefont {Shen}}, \bibinfo {author} {\bibfnamefont {H.}~\bibnamefont {Yang}}, \bibinfo {author} {\bibfnamefont {P.}~\bibnamefont {Peng}}, \bibinfo {author} {\bibfnamefont {F.}~\bibnamefont {Liu}}, \ and\ \bibinfo {author} {\bibfnamefont {Q.}~\bibnamefont {Du}},\ }\bibfield  {title} {\enquote {\bibinfo {title} {Realizable seismic cloak via polar metamaterials},}\ }\href@noop {} {\bibfield  {journal} {\bibinfo  {journal} {Appl. Phys. Lett.}\ }\textbf {\bibinfo {volume} {125}},\ \bibinfo {pages} {191701} (\bibinfo {year} {2024})}\BibitemShut {NoStop}%
\bibitem [{\citenamefont {Xu}\ \emph {et~al.}(2020)\citenamefont {Xu}, \citenamefont {Wang}, \citenamefont {Shou}, \citenamefont {Du}, \citenamefont {Chen}, \citenamefont {Li}, \citenamefont {Matusik}, \citenamefont {Hussein},\ and\ \citenamefont {Huang}}]{xu2020physical}%
  \BibitemOpen
  \bibfield  {author} {\bibinfo {author} {\bibfnamefont {X.}~\bibnamefont {Xu}}, \bibinfo {author} {\bibfnamefont {C.}~\bibnamefont {Wang}}, \bibinfo {author} {\bibfnamefont {W.}~\bibnamefont {Shou}}, \bibinfo {author} {\bibfnamefont {Z.}~\bibnamefont {Du}}, \bibinfo {author} {\bibfnamefont {Y.}~\bibnamefont {Chen}}, \bibinfo {author} {\bibfnamefont {B.}~\bibnamefont {Li}}, \bibinfo {author} {\bibfnamefont {W.}~\bibnamefont {Matusik}}, \bibinfo {author} {\bibfnamefont {N.}~\bibnamefont {Hussein}}, \ and\ \bibinfo {author} {\bibfnamefont {G.}~\bibnamefont {Huang}},\ }\bibfield  {title} {\enquote {\bibinfo {title} {Physical realization of elastic cloaking with a polar material},}\ }\href@noop {} {\bibfield  {journal} {\bibinfo  {journal} {Phys. Rev. Lett.}\ }\textbf {\bibinfo {volume} {124}},\ \bibinfo {pages} {114301} (\bibinfo {year} {2020})}\BibitemShut {NoStop}%
\bibitem [{\citenamefont {Yang}\ \emph {et~al.}(2025)\citenamefont {Yang}, \citenamefont {Yi}, \citenamefont {Li}, \citenamefont {Li}, \citenamefont {Ye}, \citenamefont {Li},\ and\ \citenamefont {Christensen}}]{yang2025static}%
  \BibitemOpen
  \bibfield  {author} {\bibinfo {author} {\bibfnamefont {Z.}~\bibnamefont {Yang}}, \bibinfo {author} {\bibfnamefont {J.}~\bibnamefont {Yi}}, \bibinfo {author} {\bibfnamefont {F.}~\bibnamefont {Li}}, \bibinfo {author} {\bibfnamefont {Z.}~\bibnamefont {Li}}, \bibinfo {author} {\bibfnamefont {L.}~\bibnamefont {Ye}}, \bibinfo {author} {\bibfnamefont {B.}~\bibnamefont {Li}}, \ and\ \bibinfo {author} {\bibfnamefont {J.}~\bibnamefont {Christensen}},\ }\bibfield  {title} {\enquote {\bibinfo {title} {Static mechanical cloaking and camouflage from disorder},}\ }\href@noop {} {\bibfield  {journal} {\bibinfo  {journal} {Nat. Commun.}\ }\textbf {\bibinfo {volume} {16}},\ \bibinfo {pages} {8858} (\bibinfo {year} {2025})}\BibitemShut {NoStop}%
\bibitem [{\citenamefont {Hai}, \citenamefont {Zhao},\ and\ \citenamefont {Meng}(2018)}]{hai2018unfeelable}%
  \BibitemOpen
  \bibfield  {author} {\bibinfo {author} {\bibfnamefont {L.}~\bibnamefont {Hai}}, \bibinfo {author} {\bibfnamefont {Q.}~\bibnamefont {Zhao}}, \ and\ \bibinfo {author} {\bibfnamefont {Y.}~\bibnamefont {Meng}},\ }\bibfield  {title} {\enquote {\bibinfo {title} {Unfeelable mechanical cloak based on proportional parameter transform in bimode structures},}\ }\href@noop {} {\bibfield  {journal} {\bibinfo  {journal} {Adv. Funct. Mater.}\ }\textbf {\bibinfo {volume} {28}},\ \bibinfo {pages} {1801473} (\bibinfo {year} {2018})}\BibitemShut {NoStop}%
\bibitem [{\citenamefont {Yavari}\ and\ \citenamefont {Golgoon}(2019)}]{yavari2019nonlinear}%
  \BibitemOpen
  \bibfield  {author} {\bibinfo {author} {\bibfnamefont {A.}~\bibnamefont {Yavari}}\ and\ \bibinfo {author} {\bibfnamefont {A.}~\bibnamefont {Golgoon}},\ }\bibfield  {title} {\enquote {\bibinfo {title} {Nonlinear and linear elastodynamic transformation cloaking},}\ }\href@noop {} {\bibfield  {journal} {\bibinfo  {journal} {Arch. Rational Mech. Anal.}\ }\textbf {\bibinfo {volume} {234}},\ \bibinfo {pages} {211} (\bibinfo {year} {2019})}\BibitemShut {NoStop}%
\bibitem [{\citenamefont {Fachinotti}, \citenamefont {Peralta},\ and\ \citenamefont {Albanesi}(2018)}]{fachinotti2018optimization}%
  \BibitemOpen
  \bibfield  {author} {\bibinfo {author} {\bibfnamefont {V.~D.}\ \bibnamefont {Fachinotti}}, \bibinfo {author} {\bibfnamefont {I.}~\bibnamefont {Peralta}}, \ and\ \bibinfo {author} {\bibfnamefont {A.~E.}\ \bibnamefont {Albanesi}},\ }\bibfield  {title} {\enquote {\bibinfo {title} {Optimization-based design of an elastostatic cloaking device},}\ }\href@noop {} {\bibfield  {journal} {\bibinfo  {journal} {Sci. Rep.}\ }\textbf {\bibinfo {volume} {8}},\ \bibinfo {pages} {9857} (\bibinfo {year} {2018})}\BibitemShut {NoStop}%
\bibitem [{\citenamefont {Sozio}, \citenamefont {Shojaei},\ and\ \citenamefont {Yavari}(2023)}]{sozio2023optimal}%
  \BibitemOpen
  \bibfield  {author} {\bibinfo {author} {\bibfnamefont {F.}~\bibnamefont {Sozio}}, \bibinfo {author} {\bibfnamefont {M.~F.}\ \bibnamefont {Shojaei}}, \ and\ \bibinfo {author} {\bibfnamefont {A.}~\bibnamefont {Yavari}},\ }\bibfield  {title} {\enquote {\bibinfo {title} {Optimal elastostatic cloaks},}\ }\href@noop {} {\bibfield  {journal} {\bibinfo  {journal} {J. Mech. Phys. Solids}\ }\textbf {\bibinfo {volume} {176}},\ \bibinfo {pages} {105306} (\bibinfo {year} {2023})}\BibitemShut {NoStop}%
\bibitem [{\citenamefont {Eshelby}(1957)}]{eshelby1957determination}%
  \BibitemOpen
  \bibfield  {author} {\bibinfo {author} {\bibfnamefont {J.~D.}\ \bibnamefont {Eshelby}},\ }\bibfield  {title} {\enquote {\bibinfo {title} {The determination of the elastic field of an ellipsoidal inclusion, and related problems},}\ }\href@noop {} {\bibfield  {journal} {\bibinfo  {journal} {Proc. R. Soc. A}\ }\textbf {\bibinfo {volume} {241}},\ \bibinfo {pages} {376} (\bibinfo {year} {1957})}\BibitemShut {NoStop}%
\bibitem [{\citenamefont {Christensen}(1993)}]{christensen1993effective}%
  \BibitemOpen
  \bibfield  {author} {\bibinfo {author} {\bibfnamefont {R.~M.}\ \bibnamefont {Christensen}},\ }\bibfield  {title} {\enquote {\bibinfo {title} {Effective properties of composite materials containing voids},}\ }\href@noop {} {\bibfield  {journal} {\bibinfo  {journal} {Proc. R. Soc. A}\ }\textbf {\bibinfo {volume} {440}},\ \bibinfo {pages} {461} (\bibinfo {year} {1993})}\BibitemShut {NoStop}%
\bibitem [{\citenamefont {Christensen}(2005)}]{Christensen2005}%
  \BibitemOpen
  \bibfield  {author} {\bibinfo {author} {\bibfnamefont {R.~M.}\ \bibnamefont {Christensen}},\ }\href@noop {} {\emph {\bibinfo {title} {Mechanics of Composite Materials}}}\ (\bibinfo  {publisher} {Dover Publications},\ \bibinfo {address} {Mineola, NY},\ \bibinfo {year} {2005})\BibitemShut {NoStop}%
\bibitem [{\citenamefont {Christensen}\ and\ \citenamefont {Lo}(1979)}]{christensen1979solutions}%
  \BibitemOpen
  \bibfield  {author} {\bibinfo {author} {\bibfnamefont {R.}~\bibnamefont {Christensen}}\ and\ \bibinfo {author} {\bibfnamefont {K.}~\bibnamefont {Lo}},\ }\bibfield  {title} {\enquote {\bibinfo {title} {Solutions for effective shear properties in three phase sphere and cylinder models},}\ }\href@noop {} {\bibfield  {journal} {\bibinfo  {journal} {J. Mech. Phys. Solids}\ }\textbf {\bibinfo {volume} {27}},\ \bibinfo {pages} {315} (\bibinfo {year} {1979})}\BibitemShut {NoStop}%
\bibitem [{\citenamefont {Hsiao-Sheng}\ and\ \citenamefont {Acrivos}(1978)}]{hsiao1978effective}%
  \BibitemOpen
  \bibfield  {author} {\bibinfo {author} {\bibfnamefont {C.}~\bibnamefont {Hsiao-Sheng}}\ and\ \bibinfo {author} {\bibfnamefont {A.}~\bibnamefont {Acrivos}},\ }\bibfield  {title} {\enquote {\bibinfo {title} {The effective elastic moduli of composite materials containing spherical inclusions at non-dilute concentrations},}\ }\href@noop {} {\bibfield  {journal} {\bibinfo  {journal} {Int. J. Solids Struct.}\ }\textbf {\bibinfo {volume} {14}},\ \bibinfo {pages} {349} (\bibinfo {year} {1978})}\BibitemShut {NoStop}%
\bibitem [{\citenamefont {Benveniste}(2008)}]{benveniste2008revisiting}%
  \BibitemOpen
  \bibfield  {author} {\bibinfo {author} {\bibfnamefont {Y.}~\bibnamefont {Benveniste}},\ }\bibfield  {title} {\enquote {\bibinfo {title} {Revisiting the generalized self-consistent scheme in composites: Clarification of some aspects and a new formulation},}\ }\href@noop {} {\bibfield  {journal} {\bibinfo  {journal} {J. Mech. Phys. Solids}\ }\textbf {\bibinfo {volume} {56}},\ \bibinfo {pages} {2984} (\bibinfo {year} {2008})}\BibitemShut {NoStop}%
\bibitem [{\citenamefont {Phan-Thien}\ and\ \citenamefont {Kim}(1994)}]{phan1994load}%
  \BibitemOpen
  \bibfield  {author} {\bibinfo {author} {\bibfnamefont {N.}~\bibnamefont {Phan-Thien}}\ and\ \bibinfo {author} {\bibfnamefont {S.}~\bibnamefont {Kim}},\ }\bibfield  {title} {\enquote {\bibinfo {title} {The load transfer between two rigid spherical inclusions in an elastic medium},}\ }\href@noop {} {\bibfield  {journal} {\bibinfo  {journal} {Z. angew. Math. Phys. ZAMP}\ }\textbf {\bibinfo {volume} {45}},\ \bibinfo {pages} {177} (\bibinfo {year} {1994})}\BibitemShut {NoStop}%
\bibitem [{\citenamefont {Puljiz}\ and\ \citenamefont {Menzel}(2019)}]{puljiz2019displacement}%
  \BibitemOpen
  \bibfield  {author} {\bibinfo {author} {\bibfnamefont {M.}~\bibnamefont {Puljiz}}\ and\ \bibinfo {author} {\bibfnamefont {A.~M.}\ \bibnamefont {Menzel}},\ }\bibfield  {title} {\enquote {\bibinfo {title} {Displacement field around a rigid sphere in a compressible elastic environment, corresponding higher-order {F}ax{\'e}n relations, as well as higher-order displaceability and rotateability matrices},}\ }\href@noop {} {\bibfield  {journal} {\bibinfo  {journal} {Phys. Rev. E}\ }\textbf {\bibinfo {volume} {99}},\ \bibinfo {pages} {053002} (\bibinfo {year} {2019})}\BibitemShut {NoStop}%
\bibitem [{\citenamefont {Sburlati}(2013)}]{sburlati2013stress}%
  \BibitemOpen
  \bibfield  {author} {\bibinfo {author} {\bibfnamefont {R.}~\bibnamefont {Sburlati}},\ }\bibfield  {title} {\enquote {\bibinfo {title} {Stress concentration factor due to a functionally graded ring around a hole in an isotropic plate},}\ }\href@noop {} {\bibfield  {journal} {\bibinfo  {journal} {Int. J. Solids Struct.}\ }\textbf {\bibinfo {volume} {50}},\ \bibinfo {pages} {3649} (\bibinfo {year} {2013})}\BibitemShut {NoStop}%
\bibitem [{\citenamefont {Sburlati}, \citenamefont {Atashipour},\ and\ \citenamefont {Atashipour}(2014)}]{sburlati2014reduction}%
  \BibitemOpen
  \bibfield  {author} {\bibinfo {author} {\bibfnamefont {R.}~\bibnamefont {Sburlati}}, \bibinfo {author} {\bibfnamefont {S.~R.}\ \bibnamefont {Atashipour}}, \ and\ \bibinfo {author} {\bibfnamefont {S.~A.}\ \bibnamefont {Atashipour}},\ }\bibfield  {title} {\enquote {\bibinfo {title} {Reduction of the stress concentration factor in a homogeneous panel with hole by using a functionally graded layer},}\ }\href@noop {} {\bibfield  {journal} {\bibinfo  {journal} {Compos. B Eng.}\ }\textbf {\bibinfo {volume} {61}},\ \bibinfo {pages} {99} (\bibinfo {year} {2014})}\BibitemShut {NoStop}%
\bibitem [{\citenamefont {Fielding}(2024)}]{fielding2024simple}%
  \BibitemOpen
  \bibfield  {author} {\bibinfo {author} {\bibfnamefont {S.~M.}\ \bibnamefont {Fielding}},\ }\bibfield  {title} {\enquote {\bibinfo {title} {Simple and effective mechanical cloaking},}\ }\href@noop {} {\bibfield  {journal} {\bibinfo  {journal} {J. Mech. Phys. Solids}\ }\textbf {\bibinfo {volume} {192}},\ \bibinfo {pages} {105824} (\bibinfo {year} {2024})}\BibitemShut {NoStop}%
\bibitem [{\citenamefont {Herlach}\ \emph {et~al.}(2010)\citenamefont {Herlach}, \citenamefont {Klassen}, \citenamefont {Wette},\ and\ \citenamefont {Holland-Moritz}}]{herlach2010colloids}%
  \BibitemOpen
  \bibfield  {author} {\bibinfo {author} {\bibfnamefont {D.~M.}\ \bibnamefont {Herlach}}, \bibinfo {author} {\bibfnamefont {I.}~\bibnamefont {Klassen}}, \bibinfo {author} {\bibfnamefont {P.}~\bibnamefont {Wette}}, \ and\ \bibinfo {author} {\bibfnamefont {D.}~\bibnamefont {Holland-Moritz}},\ }\bibfield  {title} {\enquote {\bibinfo {title} {Colloids as model systems for metals and alloys: a case study of crystallization},}\ }\href@noop {} {\bibfield  {journal} {\bibinfo  {journal} {J. Phys.: Condens. Matter}\ }\textbf {\bibinfo {volume} {22}},\ \bibinfo {pages} {153101} (\bibinfo {year} {2010})}\BibitemShut {NoStop}%
\bibitem [{\citenamefont {Royall}\ \emph {et~al.}(2024)\citenamefont {Royall}, \citenamefont {Charbonneau}, \citenamefont {Dijkstra}, \citenamefont {Russo}, \citenamefont {Smallenburg}, \citenamefont {Speck},\ and\ \citenamefont {Valeriani}}]{royall2024colloidal}%
  \BibitemOpen
  \bibfield  {author} {\bibinfo {author} {\bibfnamefont {C.~P.}\ \bibnamefont {Royall}}, \bibinfo {author} {\bibfnamefont {P.}~\bibnamefont {Charbonneau}}, \bibinfo {author} {\bibfnamefont {M.}~\bibnamefont {Dijkstra}}, \bibinfo {author} {\bibfnamefont {J.}~\bibnamefont {Russo}}, \bibinfo {author} {\bibfnamefont {F.}~\bibnamefont {Smallenburg}}, \bibinfo {author} {\bibfnamefont {T.}~\bibnamefont {Speck}}, \ and\ \bibinfo {author} {\bibfnamefont {C.}~\bibnamefont {Valeriani}},\ }\bibfield  {title} {\enquote {\bibinfo {title} {Colloidal hard spheres: Triumphs, challenges, and mysteries},}\ }\href@noop {} {\bibfield  {journal} {\bibinfo  {journal} {Rev. Mod. Phys.}\ }\textbf {\bibinfo {volume} {96}},\ \bibinfo {pages} {045003} (\bibinfo {year} {2024})}\BibitemShut {NoStop}%
\bibitem [{\citenamefont {Menath}\ \emph {et~al.}(2023)\citenamefont {Menath}, \citenamefont {Mohammadi}, \citenamefont {Grauer}, \citenamefont {Deters}, \citenamefont {B{\"o}hm}, \citenamefont {Liebchen}, \citenamefont {Janssen}, \citenamefont {L{\"o}wen},\ and\ \citenamefont {Vogel}}]{menath2023acoustic}%
  \BibitemOpen
  \bibfield  {author} {\bibinfo {author} {\bibfnamefont {J.}~\bibnamefont {Menath}}, \bibinfo {author} {\bibfnamefont {R.}~\bibnamefont {Mohammadi}}, \bibinfo {author} {\bibfnamefont {J.~C.}\ \bibnamefont {Grauer}}, \bibinfo {author} {\bibfnamefont {C.}~\bibnamefont {Deters}}, \bibinfo {author} {\bibfnamefont {M.}~\bibnamefont {B{\"o}hm}}, \bibinfo {author} {\bibfnamefont {B.}~\bibnamefont {Liebchen}}, \bibinfo {author} {\bibfnamefont {L.~M.}\ \bibnamefont {Janssen}}, \bibinfo {author} {\bibfnamefont {H.}~\bibnamefont {L{\"o}wen}}, \ and\ \bibinfo {author} {\bibfnamefont {N.}~\bibnamefont {Vogel}},\ }\bibfield  {title} {\enquote {\bibinfo {title} {Acoustic crystallization of 2{D} colloidal crystals},}\ }\href@noop {} {\bibfield  {journal} {\bibinfo  {journal} {Adv. Mater.}\ }\textbf {\bibinfo {volume} {35}},\ \bibinfo {pages} {2206593} (\bibinfo {year} {2023})}\BibitemShut {NoStop}%
\bibitem [{\citenamefont {Landau}\ and\ \citenamefont {Lifshitz}(1999)}]{landau2012theory}%
  \BibitemOpen
  \bibfield  {author} {\bibinfo {author} {\bibfnamefont {L.~D.}\ \bibnamefont {Landau}}\ and\ \bibinfo {author} {\bibfnamefont {E.~M.}\ \bibnamefont {Lifshitz}},\ }\href@noop {} {\emph {\bibinfo {title} {{Theory of Elasticity}}}}\ (\bibinfo  {publisher} {Butterworth-Heinemann, Oxford},\ \bibinfo {year} {1999})\BibitemShut {NoStop}%
\bibitem [{\citenamefont {Chaikin}\ and\ \citenamefont {Lubensky}(1995)}]{chaikin1995principles}%
  \BibitemOpen
  \bibfield  {author} {\bibinfo {author} {\bibfnamefont {P.~M.}\ \bibnamefont {Chaikin}}\ and\ \bibinfo {author} {\bibfnamefont {T.~C.}\ \bibnamefont {Lubensky}},\ }\href@noop {} {\emph {\bibinfo {title} {Principles of condensed matter physics}}}\ (\bibinfo  {publisher} {Cambridge University Press, New York},\ \bibinfo {year} {1995})\BibitemShut {NoStop}%
\bibitem [{\citenamefont {Michell}(1899)}]{michell1899direct}%
  \BibitemOpen
  \bibfield  {author} {\bibinfo {author} {\bibfnamefont {J.~H.}\ \bibnamefont {Michell}},\ }\bibfield  {title} {\enquote {\bibinfo {title} {On the direct determination of stress in an elastic solid, with application to the theory of plates},}\ }\href@noop {} {\bibfield  {journal} {\bibinfo  {journal} {Proc. London Math. Soc.}\ }\textbf {\bibinfo {volume} {1}},\ \bibinfo {pages} {100} (\bibinfo {year} {1899})}\BibitemShut {NoStop}%
\bibitem [{\citenamefont {Greaves}\ \emph {et~al.}(2011)\citenamefont {Greaves}, \citenamefont {Greer}, \citenamefont {Lakes},\ and\ \citenamefont {Rouxel}}]{greaves2011poisson}%
  \BibitemOpen
  \bibfield  {author} {\bibinfo {author} {\bibfnamefont {G.~N.}\ \bibnamefont {Greaves}}, \bibinfo {author} {\bibfnamefont {A.~L.}\ \bibnamefont {Greer}}, \bibinfo {author} {\bibfnamefont {R.~S.}\ \bibnamefont {Lakes}}, \ and\ \bibinfo {author} {\bibfnamefont {T.}~\bibnamefont {Rouxel}},\ }\bibfield  {title} {\enquote {\bibinfo {title} {Poisson's ratio and modern materials},}\ }\href@noop {} {\bibfield  {journal} {\bibinfo  {journal} {Nat. Mater.}\ }\textbf {\bibinfo {volume} {10}},\ \bibinfo {pages} {823} (\bibinfo {year} {2011})}\BibitemShut {NoStop}%
\bibitem [{\citenamefont {Mott}\ and\ \citenamefont {Roland}(2009)}]{mott2009limits}%
  \BibitemOpen
  \bibfield  {author} {\bibinfo {author} {\bibfnamefont {P.}~\bibnamefont {Mott}}\ and\ \bibinfo {author} {\bibfnamefont {C.}~\bibnamefont {Roland}},\ }\bibfield  {title} {\enquote {\bibinfo {title} {Limits to {P}oisson’s ratio in isotropic materials},}\ }\href@noop {} {\bibfield  {journal} {\bibinfo  {journal} {Phys. Rev. B}\ }\textbf {\bibinfo {volume} {80}},\ \bibinfo {pages} {132104} (\bibinfo {year} {2009})}\BibitemShut {NoStop}%
\bibitem [{\citenamefont {Soman}\ \emph {et~al.}(2012)\citenamefont {Soman}, \citenamefont {Fozdar}, \citenamefont {Lee}, \citenamefont {Phadke}, \citenamefont {Varghese},\ and\ \citenamefont {Chen}}]{soman2012three}%
  \BibitemOpen
  \bibfield  {author} {\bibinfo {author} {\bibfnamefont {P.}~\bibnamefont {Soman}}, \bibinfo {author} {\bibfnamefont {D.~Y.}\ \bibnamefont {Fozdar}}, \bibinfo {author} {\bibfnamefont {J.~W.}\ \bibnamefont {Lee}}, \bibinfo {author} {\bibfnamefont {A.}~\bibnamefont {Phadke}}, \bibinfo {author} {\bibfnamefont {S.}~\bibnamefont {Varghese}}, \ and\ \bibinfo {author} {\bibfnamefont {S.}~\bibnamefont {Chen}},\ }\bibfield  {title} {\enquote {\bibinfo {title} {A three-dimensional polymer scaffolding material exhibiting a zero {P}oisson's ratio},}\ }\href@noop {} {\bibfield  {journal} {\bibinfo  {journal} {Soft Matter}\ }\textbf {\bibinfo {volume} {8}},\ \bibinfo {pages} {4946} (\bibinfo {year} {2012})}\BibitemShut {NoStop}%
\bibitem [{\citenamefont {Jiang}\ and\ \citenamefont {Park}(2014)}]{jiang2014negative}%
  \BibitemOpen
  \bibfield  {author} {\bibinfo {author} {\bibfnamefont {J.-W.}\ \bibnamefont {Jiang}}\ and\ \bibinfo {author} {\bibfnamefont {H.~S.}\ \bibnamefont {Park}},\ }\bibfield  {title} {\enquote {\bibinfo {title} {Negative {P}oisson’s ratio in single-layer black phosphorus},}\ }\href@noop {} {\bibfield  {journal} {\bibinfo  {journal} {Nat. Commun.}\ }\textbf {\bibinfo {volume} {5}},\ \bibinfo {pages} {4727} (\bibinfo {year} {2014})}\BibitemShut {NoStop}%
\bibitem [{\citenamefont {Yu}, \citenamefont {Yan},\ and\ \citenamefont {Ruzsinszky}(2017)}]{yu2017negative}%
  \BibitemOpen
  \bibfield  {author} {\bibinfo {author} {\bibfnamefont {L.}~\bibnamefont {Yu}}, \bibinfo {author} {\bibfnamefont {Q.}~\bibnamefont {Yan}}, \ and\ \bibinfo {author} {\bibfnamefont {A.}~\bibnamefont {Ruzsinszky}},\ }\bibfield  {title} {\enquote {\bibinfo {title} {Negative {P}oisson’s ratio in 1\uppercase{T}-type crystalline two-dimensional transition metal dichalcogenides},}\ }\href@noop {} {\bibfield  {journal} {\bibinfo  {journal} {Nat. Commun.}\ }\textbf {\bibinfo {volume} {8}},\ \bibinfo {pages} {15224} (\bibinfo {year} {2017})}\BibitemShut {NoStop}%
\bibitem [{\citenamefont {Liu}\ and\ \citenamefont {Zhang}(2018)}]{liu2018soft}%
  \BibitemOpen
  \bibfield  {author} {\bibinfo {author} {\bibfnamefont {J.}~\bibnamefont {Liu}}\ and\ \bibinfo {author} {\bibfnamefont {Y.}~\bibnamefont {Zhang}},\ }\bibfield  {title} {\enquote {\bibinfo {title} {Soft network materials with isotropic negative {P}oisson's ratios over large strains},}\ }\href@noop {} {\bibfield  {journal} {\bibinfo  {journal} {Soft Matter}\ }\textbf {\bibinfo {volume} {14}},\ \bibinfo {pages} {693} (\bibinfo {year} {2018})}\BibitemShut {NoStop}%
\bibitem [{\citenamefont {Lee}\ \emph {et~al.}(2019)\citenamefont {Lee}, \citenamefont {Lim}, \citenamefont {Yi}, \citenamefont {Lee}, \citenamefont {Kang}, \citenamefont {Choi}, \citenamefont {Han}, \citenamefont {Sun}, \citenamefont {Choi},\ and\ \citenamefont {Joo}}]{lee2019auxetic}%
  \BibitemOpen
  \bibfield  {author} {\bibinfo {author} {\bibfnamefont {Y.-J.}\ \bibnamefont {Lee}}, \bibinfo {author} {\bibfnamefont {S.-M.}\ \bibnamefont {Lim}}, \bibinfo {author} {\bibfnamefont {S.-M.}\ \bibnamefont {Yi}}, \bibinfo {author} {\bibfnamefont {J.-H.}\ \bibnamefont {Lee}}, \bibinfo {author} {\bibfnamefont {S.-g.}\ \bibnamefont {Kang}}, \bibinfo {author} {\bibfnamefont {G.-M.}\ \bibnamefont {Choi}}, \bibinfo {author} {\bibfnamefont {H.~N.}\ \bibnamefont {Han}}, \bibinfo {author} {\bibfnamefont {J.-Y.}\ \bibnamefont {Sun}}, \bibinfo {author} {\bibfnamefont {I.-S.}\ \bibnamefont {Choi}}, \ and\ \bibinfo {author} {\bibfnamefont {Y.-C.}\ \bibnamefont {Joo}},\ }\bibfield  {title} {\enquote {\bibinfo {title} {Auxetic elastomers: Mechanically programmable meta-elastomers with an unusual {P}oisson’s ratio overcome the gauge limit of a capacitive type strain sensor},}\ }\href@noop {} {\bibfield  {journal} {\bibinfo  {journal} {Extrem. Mech. Lett.}\ }\textbf {\bibinfo {volume} {31}},\ \bibinfo {pages} {100516}
  (\bibinfo {year} {2019})}\BibitemShut {NoStop}%
\bibitem [{\citenamefont {Gaal}\ \emph {et~al.}(2020)\citenamefont {Gaal}, \citenamefont {Rodrigues}, \citenamefont {Dantas}, \citenamefont {Galv{\~a}o},\ and\ \citenamefont {Fonseca}}]{gaal2020new}%
  \BibitemOpen
  \bibfield  {author} {\bibinfo {author} {\bibfnamefont {V.}~\bibnamefont {Gaal}}, \bibinfo {author} {\bibfnamefont {V.}~\bibnamefont {Rodrigues}}, \bibinfo {author} {\bibfnamefont {S.~O.}\ \bibnamefont {Dantas}}, \bibinfo {author} {\bibfnamefont {D.~S.}\ \bibnamefont {Galv{\~a}o}}, \ and\ \bibinfo {author} {\bibfnamefont {A.~F.}\ \bibnamefont {Fonseca}},\ }\bibfield  {title} {\enquote {\bibinfo {title} {New zero {P}oisson's ratio structures},}\ }\href@noop {} {\bibfield  {journal} {\bibinfo  {journal} {Phys. Status Solidi - Rapid Res. Lett.}\ }\textbf {\bibinfo {volume} {14}},\ \bibinfo {pages} {1900564} (\bibinfo {year} {2020})}\BibitemShut {NoStop}%
\bibitem [{\citenamefont {Gao}\ \emph {et~al.}(2024)\citenamefont {Gao}, \citenamefont {Shi}, \citenamefont {Qiao}, \citenamefont {Yu},\ and\ \citenamefont {Yin}}]{gao2024strain}%
  \BibitemOpen
  \bibfield  {author} {\bibinfo {author} {\bibfnamefont {W.}~\bibnamefont {Gao}}, \bibinfo {author} {\bibfnamefont {X.}~\bibnamefont {Shi}}, \bibinfo {author} {\bibfnamefont {Y.}~\bibnamefont {Qiao}}, \bibinfo {author} {\bibfnamefont {M.}~\bibnamefont {Yu}}, \ and\ \bibinfo {author} {\bibfnamefont {H.}~\bibnamefont {Yin}},\ }\bibfield  {title} {\enquote {\bibinfo {title} {Strain-dependent near-zero and negative {P}oisson ratios in a two-dimensional (\uppercase{C}u\uppercase{I})\uppercase{P}$_4$\uppercase{S}e$_4$ monolayer},}\ }\href@noop {} {\bibfield  {journal} {\bibinfo  {journal} {Phys. Rev. B}\ }\textbf {\bibinfo {volume} {109}},\ \bibinfo {pages} {075402} (\bibinfo {year} {2024})}\BibitemShut {NoStop}%
\bibitem [{\citenamefont {Grima}\ \emph {et~al.}(2015)\citenamefont {Grima}, \citenamefont {Winczewski}, \citenamefont {Mizzi}, \citenamefont {Grech}, \citenamefont {Cauchi}, \citenamefont {Gatt}, \citenamefont {Attard}, \citenamefont {Wojciechowski},\ and\ \citenamefont {Rybicki}}]{grima2015tailoring}%
  \BibitemOpen
  \bibfield  {author} {\bibinfo {author} {\bibfnamefont {J.~N.}\ \bibnamefont {Grima}}, \bibinfo {author} {\bibfnamefont {S.}~\bibnamefont {Winczewski}}, \bibinfo {author} {\bibfnamefont {L.}~\bibnamefont {Mizzi}}, \bibinfo {author} {\bibfnamefont {M.~C.}\ \bibnamefont {Grech}}, \bibinfo {author} {\bibfnamefont {R.}~\bibnamefont {Cauchi}}, \bibinfo {author} {\bibfnamefont {R.}~\bibnamefont {Gatt}}, \bibinfo {author} {\bibfnamefont {D.}~\bibnamefont {Attard}}, \bibinfo {author} {\bibfnamefont {K.~W.}\ \bibnamefont {Wojciechowski}}, \ and\ \bibinfo {author} {\bibfnamefont {J.}~\bibnamefont {Rybicki}},\ }\bibfield  {title} {\enquote {\bibinfo {title} {Tailoring graphene to achieve negative {P}oisson's ratio properties},}\ }\href@noop {} {\bibfield  {journal} {\bibinfo  {journal} {Adv. Mater.}\ }\textbf {\bibinfo {volume} {27}},\ \bibinfo {pages} {1455} (\bibinfo {year} {2015})}\BibitemShut {NoStop}%
\bibitem [{\citenamefont {Grima}\ \emph {et~al.}(2018)\citenamefont {Grima}, \citenamefont {Grech}, \citenamefont {Grima-Cornish}, \citenamefont {Gatt},\ and\ \citenamefont {Attard}}]{grima2018giant}%
  \BibitemOpen
  \bibfield  {author} {\bibinfo {author} {\bibfnamefont {J.~N.}\ \bibnamefont {Grima}}, \bibinfo {author} {\bibfnamefont {M.~C.}\ \bibnamefont {Grech}}, \bibinfo {author} {\bibfnamefont {J.~N.}\ \bibnamefont {Grima-Cornish}}, \bibinfo {author} {\bibfnamefont {R.}~\bibnamefont {Gatt}}, \ and\ \bibinfo {author} {\bibfnamefont {D.}~\bibnamefont {Attard}},\ }\bibfield  {title} {\enquote {\bibinfo {title} {Giant auxetic behaviour in engineered graphene},}\ }\href@noop {} {\bibfield  {journal} {\bibinfo  {journal} {Annalen der Physik}\ }\textbf {\bibinfo {volume} {530}},\ \bibinfo {pages} {1700330} (\bibinfo {year} {2018})}\BibitemShut {NoStop}%
\bibitem [{\citenamefont {Toxvaerd}\ and\ \citenamefont {Dyre}(2011)}]{toxvaerd2011communication}%
  \BibitemOpen
  \bibfield  {author} {\bibinfo {author} {\bibfnamefont {S.}~\bibnamefont {Toxvaerd}}\ and\ \bibinfo {author} {\bibfnamefont {J.~C.}\ \bibnamefont {Dyre}},\ }\bibfield  {title} {\enquote {\bibinfo {title} {Communication: Shifted forces in molecular dynamics},}\ }\href@noop {} {\bibfield  {journal} {\bibinfo  {journal} {J. Chem. Phys.}\ }\textbf {\bibinfo {volume} {134}},\ \bibinfo {pages} {081102} (\bibinfo {year} {2011})}\BibitemShut {NoStop}%
\bibitem [{\citenamefont {Thompson}\ \emph {et~al.}(2022)\citenamefont {Thompson}, \citenamefont {Aktulga}, \citenamefont {Berger}, \citenamefont {Bolintineanu}, \citenamefont {Brown}, \citenamefont {Crozier}, \citenamefont {in't Veld}, \citenamefont {Kohlmeyer}, \citenamefont {Moore}, \citenamefont {Nguyen} \emph {et~al.}}]{thompson2022lammps}%
  \BibitemOpen
  \bibfield  {author} {\bibinfo {author} {\bibfnamefont {A.~P.}\ \bibnamefont {Thompson}}, \bibinfo {author} {\bibfnamefont {H.~M.}\ \bibnamefont {Aktulga}}, \bibinfo {author} {\bibfnamefont {R.}~\bibnamefont {Berger}}, \bibinfo {author} {\bibfnamefont {D.~S.}\ \bibnamefont {Bolintineanu}}, \bibinfo {author} {\bibfnamefont {W.~M.}\ \bibnamefont {Brown}}, \bibinfo {author} {\bibfnamefont {P.~S.}\ \bibnamefont {Crozier}}, \bibinfo {author} {\bibfnamefont {P.~J.}\ \bibnamefont {in't Veld}}, \bibinfo {author} {\bibfnamefont {A.}~\bibnamefont {Kohlmeyer}}, \bibinfo {author} {\bibfnamefont {S.~G.}\ \bibnamefont {Moore}}, \bibinfo {author} {\bibfnamefont {T.~D.}\ \bibnamefont {Nguyen}},  \emph {et~al.},\ }\bibfield  {title} {\enquote {\bibinfo {title} {{LAMMPS} -- a flexible simulation tool for particle-based materials modeling at the atomic, meso, and continuum scales},}\ }\href@noop {} {\bibfield  {journal} {\bibinfo  {journal} {Comput. Phys. Commun.}\ }\textbf {\bibinfo {volume} {271}},\ \bibinfo {pages} {108171}
  (\bibinfo {year} {2022})}\BibitemShut {NoStop}%
\bibitem [{Lam()}]{Lammps}%
  \BibitemOpen
  \href@noop {} {\enquote {\bibinfo {title} {{LAMMPS} software package},}\ }\bibinfo {howpublished} {\url{https://www.lammps.org}}\BibitemShut {NoStop}%
\bibitem [{\citenamefont {Fisher}, \citenamefont {Halperin},\ and\ \citenamefont {Morf}(1979)}]{fisher1979defects}%
  \BibitemOpen
  \bibfield  {author} {\bibinfo {author} {\bibfnamefont {D.~S.}\ \bibnamefont {Fisher}}, \bibinfo {author} {\bibfnamefont {B.}~\bibnamefont {Halperin}}, \ and\ \bibinfo {author} {\bibfnamefont {R.}~\bibnamefont {Morf}},\ }\bibfield  {title} {\enquote {\bibinfo {title} {Defects in the two-dimensional electron solid and implications for melting},}\ }\href@noop {} {\bibfield  {journal} {\bibinfo  {journal} {Phys. Rev. B}\ }\textbf {\bibinfo {volume} {20}},\ \bibinfo {pages} {4692} (\bibinfo {year} {1979})}\BibitemShut {NoStop}%
\bibitem [{\citenamefont {Zhang}\ and\ \citenamefont {Nelson}(2021)}]{zhang2021phonon}%
  \BibitemOpen
  \bibfield  {author} {\bibinfo {author} {\bibfnamefont {G.~H.}\ \bibnamefont {Zhang}}\ and\ \bibinfo {author} {\bibfnamefont {D.~R.}\ \bibnamefont {Nelson}},\ }\bibfield  {title} {\enquote {\bibinfo {title} {Phonon eigenfunctions of inhomogeneous lattices: Can you hear the shape of a cone?}}\ }\href@noop {} {\bibfield  {journal} {\bibinfo  {journal} {Phys. Rev. E}\ }\textbf {\bibinfo {volume} {104}},\ \bibinfo {pages} {065005} (\bibinfo {year} {2021})}\BibitemShut {NoStop}%
\bibitem [{\citenamefont {Stukowski}(2009)}]{stukowski2009visualization}%
  \BibitemOpen
  \bibfield  {author} {\bibinfo {author} {\bibfnamefont {A.}~\bibnamefont {Stukowski}},\ }\bibfield  {title} {\enquote {\bibinfo {title} {Visualization and analysis of atomistic simulation data with ovito--the open visualization tool},}\ }\href@noop {} {\bibfield  {journal} {\bibinfo  {journal} {Modell. Simul. Mater. Sci. Eng.}\ }\textbf {\bibinfo {volume} {18}},\ \bibinfo {pages} {015012} (\bibinfo {year} {2009})}\BibitemShut {NoStop}%
\bibitem [{\citenamefont {Wojciechowski}\ and\ \citenamefont {Tretiakov}(1999)}]{wojciechowski1999computer}%
  \BibitemOpen
  \bibfield  {author} {\bibinfo {author} {\bibfnamefont {K.~W.}\ \bibnamefont {Wojciechowski}}\ and\ \bibinfo {author} {\bibfnamefont {K.~V.}\ \bibnamefont {Tretiakov}},\ }\bibfield  {title} {\enquote {\bibinfo {title} {Computer simulation of elastic properties of solids under pressure},}\ }\href@noop {} {\bibfield  {journal} {\bibinfo  {journal} {Comput. Phys. Commun.}\ }\textbf {\bibinfo {volume} {121}},\ \bibinfo {pages} {528} (\bibinfo {year} {1999})}\BibitemShut {NoStop}%
\bibitem [{\citenamefont {Rechtsman}, \citenamefont {Stillinger},\ and\ \citenamefont {Torquato}(2008)}]{rechtsman2008negative}%
  \BibitemOpen
  \bibfield  {author} {\bibinfo {author} {\bibfnamefont {M.~C.}\ \bibnamefont {Rechtsman}}, \bibinfo {author} {\bibfnamefont {F.~H.}\ \bibnamefont {Stillinger}}, \ and\ \bibinfo {author} {\bibfnamefont {S.}~\bibnamefont {Torquato}},\ }\bibfield  {title} {\enquote {\bibinfo {title} {Negative {P}oisson’s ratio materials via isotropic interactions},}\ }\href@noop {} {\bibfield  {journal} {\bibinfo  {journal} {Phys. Rev. Lett.}\ }\textbf {\bibinfo {volume} {101}},\ \bibinfo {pages} {085501} (\bibinfo {year} {2008})}\BibitemShut {NoStop}%
\bibitem [{\citenamefont {Wojciechowski}(2003)}]{kw2003remarks}%
  \BibitemOpen
  \bibfield  {author} {\bibinfo {author} {\bibfnamefont {K.~W.}\ \bibnamefont {Wojciechowski}},\ }\bibfield  {title} {\enquote {\bibinfo {title} {Remarks on ``{P}oisson ratio beyond the limits of the elasticity theory''},}\ }\href@noop {} {\bibfield  {journal} {\bibinfo  {journal} {J. Phys. Soc. Jpn.}\ }\textbf {\bibinfo {volume} {72}},\ \bibinfo {pages} {1819} (\bibinfo {year} {2003})}\BibitemShut {NoStop}%
\bibitem [{\citenamefont {Gao}\ \emph {et~al.}(2021)\citenamefont {Gao}, \citenamefont {Li}, \citenamefont {Fang}, \citenamefont {Shao},\ and\ \citenamefont {Baughman}}]{gao2021bounds}%
  \BibitemOpen
  \bibfield  {author} {\bibinfo {author} {\bibfnamefont {E.}~\bibnamefont {Gao}}, \bibinfo {author} {\bibfnamefont {R.}~\bibnamefont {Li}}, \bibinfo {author} {\bibfnamefont {S.}~\bibnamefont {Fang}}, \bibinfo {author} {\bibfnamefont {Q.}~\bibnamefont {Shao}}, \ and\ \bibinfo {author} {\bibfnamefont {R.~H.}\ \bibnamefont {Baughman}},\ }\bibfield  {title} {\enquote {\bibinfo {title} {Bounds on the in-plane {P}oisson's ratios and the in-plane linear and area compressibilities for sheet crystals},}\ }\href@noop {} {\bibfield  {journal} {\bibinfo  {journal} {J. Mech. Phys. Solids}\ }\textbf {\bibinfo {volume} {152}},\ \bibinfo {pages} {104409} (\bibinfo {year} {2021})}\BibitemShut {NoStop}%
\bibitem [{\citenamefont {Eischen}\ and\ \citenamefont {Torquato}(1993)}]{eischen1993determining}%
  \BibitemOpen
  \bibfield  {author} {\bibinfo {author} {\bibfnamefont {J.}~\bibnamefont {Eischen}}\ and\ \bibinfo {author} {\bibfnamefont {S.}~\bibnamefont {Torquato}},\ }\bibfield  {title} {\enquote {\bibinfo {title} {Determining elastic behavior of composites by the boundary element method},}\ }\href@noop {} {\bibfield  {journal} {\bibinfo  {journal} {J. Appl. Phys.}\ }\textbf {\bibinfo {volume} {74}},\ \bibinfo {pages} {159} (\bibinfo {year} {1993})}\BibitemShut {NoStop}%
\bibitem [{\citenamefont {Meille}\ and\ \citenamefont {Garboczi}(2001)}]{meille2001linear}%
  \BibitemOpen
  \bibfield  {author} {\bibinfo {author} {\bibfnamefont {S.}~\bibnamefont {Meille}}\ and\ \bibinfo {author} {\bibfnamefont {E.~J.}\ \bibnamefont {Garboczi}},\ }\bibfield  {title} {\enquote {\bibinfo {title} {Linear elastic properties of 2{D} and 3{D} models of porous materialsmade from elongated objects},}\ }\href@noop {} {\bibfield  {journal} {\bibinfo  {journal} {Modell. Simul. Mater. Sci. Eng.}\ }\textbf {\bibinfo {volume} {9}},\ \bibinfo {pages} {371} (\bibinfo {year} {2001})}\BibitemShut {NoStop}%
\bibitem [{\citenamefont {Thompson}, \citenamefont {Plimpton},\ and\ \citenamefont {Mattson}(2009)}]{thompson2009general}%
  \BibitemOpen
  \bibfield  {author} {\bibinfo {author} {\bibfnamefont {A.~P.}\ \bibnamefont {Thompson}}, \bibinfo {author} {\bibfnamefont {S.~J.}\ \bibnamefont {Plimpton}}, \ and\ \bibinfo {author} {\bibfnamefont {W.}~\bibnamefont {Mattson}},\ }\bibfield  {title} {\enquote {\bibinfo {title} {General formulation of pressure and stress tensor for arbitrary many-body interaction potentials under periodic boundary conditions},}\ }\href@noop {} {\bibfield  {journal} {\bibinfo  {journal} {J. Chem. Phys.}\ }\textbf {\bibinfo {volume} {131}},\ \bibinfo {pages} {154107} (\bibinfo {year} {2009})}\BibitemShut {NoStop}%
\bibitem [{\citenamefont {Lerner}(2022)}]{lerner2022ultrahigh}%
  \BibitemOpen
  \bibfield  {author} {\bibinfo {author} {\bibfnamefont {E.}~\bibnamefont {Lerner}},\ }\bibfield  {title} {\enquote {\bibinfo {title} {Ultrahigh {P}oisson's ratio glasses},}\ }\href@noop {} {\bibfield  {journal} {\bibinfo  {journal} {Phys. Rev. Mater.}\ }\textbf {\bibinfo {volume} {6}},\ \bibinfo {pages} {065604} (\bibinfo {year} {2022})}\BibitemShut {NoStop}%
\bibitem [{\citenamefont {Jacobson}\ and\ \citenamefont {Thompson}(2022{\natexlab{a}})}]{jacobson2022revisting}%
  \BibitemOpen
  \bibfield  {author} {\bibinfo {author} {\bibfnamefont {D.~W.}\ \bibnamefont {Jacobson}}\ and\ \bibinfo {author} {\bibfnamefont {G.~B.}\ \bibnamefont {Thompson}},\ }\bibfield  {title} {\enquote {\bibinfo {title} {Revisting \uppercase{L}ennard \uppercase{J}ones, \uppercase{M}orse, and \uppercase{N}-\uppercase{M} potentials for metals},}\ }\href@noop {} {\bibfield  {journal} {\bibinfo  {journal} {Comput. Mater. Sci.}\ }\textbf {\bibinfo {volume} {205}},\ \bibinfo {pages} {111206} (\bibinfo {year} {2022}{\natexlab{a}})}\BibitemShut {NoStop}%
\bibitem [{\citenamefont {Jacobson}\ and\ \citenamefont {Thompson}(2022{\natexlab{b}})}]{jacobson2022corrigendum}%
  \BibitemOpen
  \bibfield  {author} {\bibinfo {author} {\bibfnamefont {D.}~\bibnamefont {Jacobson}}\ and\ \bibinfo {author} {\bibfnamefont {G.}~\bibnamefont {Thompson}},\ }\bibfield  {title} {\enquote {\bibinfo {title} {Corrigendum to “revisting \uppercase{L}ennard \uppercase{J}ones, \uppercase{M}orse, and \uppercase{N}-\uppercase{M} potentials for metals”[comput. mater. sci. 205, 111206 (2022)]},}\ }\href@noop {} {\bibfield  {journal} {\bibinfo  {journal} {Comput. Mater. Sci.}\ }\textbf {\bibinfo {volume} {208}},\ \bibinfo {pages} {111338} (\bibinfo {year} {2022}{\natexlab{b}})}\BibitemShut {NoStop}%
\bibitem [{\citenamefont {Filippova}, \citenamefont {Kunavin},\ and\ \citenamefont {Pugachev}(2015)}]{filippova2015calculation}%
  \BibitemOpen
  \bibfield  {author} {\bibinfo {author} {\bibfnamefont {V.}~\bibnamefont {Filippova}}, \bibinfo {author} {\bibfnamefont {S.}~\bibnamefont {Kunavin}}, \ and\ \bibinfo {author} {\bibfnamefont {M.~S.}\ \bibnamefont {Pugachev}},\ }\bibfield  {title} {\enquote {\bibinfo {title} {Calculation of the parameters of the \uppercase{L}ennard-\uppercase{J}ones potential for pairs of identical atoms based on the properties of solid substances},}\ }\href@noop {} {\bibfield  {journal} {\bibinfo  {journal} {Inorg. Mater. Appl. Res.}\ }\textbf {\bibinfo {volume} {6}},\ \bibinfo {pages} {1} (\bibinfo {year} {2015})}\BibitemShut {NoStop}%
\bibitem [{\citenamefont {Momeni}\ \emph {et~al.}(2020)\citenamefont {Momeni}, \citenamefont {Ji}, \citenamefont {Wang}, \citenamefont {Paul}, \citenamefont {Neshani}, \citenamefont {Yilmaz}, \citenamefont {Shin}, \citenamefont {Zhang}, \citenamefont {Jiang}, \citenamefont {Park} \emph {et~al.}}]{momeni2020multiscale}%
  \BibitemOpen
  \bibfield  {author} {\bibinfo {author} {\bibfnamefont {K.}~\bibnamefont {Momeni}}, \bibinfo {author} {\bibfnamefont {Y.}~\bibnamefont {Ji}}, \bibinfo {author} {\bibfnamefont {Y.}~\bibnamefont {Wang}}, \bibinfo {author} {\bibfnamefont {S.}~\bibnamefont {Paul}}, \bibinfo {author} {\bibfnamefont {S.}~\bibnamefont {Neshani}}, \bibinfo {author} {\bibfnamefont {D.~E.}\ \bibnamefont {Yilmaz}}, \bibinfo {author} {\bibfnamefont {Y.~K.}\ \bibnamefont {Shin}}, \bibinfo {author} {\bibfnamefont {D.}~\bibnamefont {Zhang}}, \bibinfo {author} {\bibfnamefont {J.-W.}\ \bibnamefont {Jiang}}, \bibinfo {author} {\bibfnamefont {H.~S.}\ \bibnamefont {Park}},  \emph {et~al.},\ }\bibfield  {title} {\enquote {\bibinfo {title} {Multiscale computational understanding and growth of 2\uppercase{D} materials: a review},}\ }\href@noop {} {\bibfield  {journal} {\bibinfo  {journal} {npj Comput. Mater.}\ }\textbf {\bibinfo {volume} {6}},\ \bibinfo {pages} {22} (\bibinfo {year} {2020})}\BibitemShut {NoStop}%
\bibitem [{\citenamefont {Rigelesaiyin}\ \emph {et~al.}(2018)\citenamefont {Rigelesaiyin}, \citenamefont {Diaz}, \citenamefont {Li}, \citenamefont {Xiong},\ and\ \citenamefont {Chen}}]{rigelesaiyin2018asymmetry}%
  \BibitemOpen
  \bibfield  {author} {\bibinfo {author} {\bibfnamefont {J.}~\bibnamefont {Rigelesaiyin}}, \bibinfo {author} {\bibfnamefont {A.}~\bibnamefont {Diaz}}, \bibinfo {author} {\bibfnamefont {W.}~\bibnamefont {Li}}, \bibinfo {author} {\bibfnamefont {L.}~\bibnamefont {Xiong}}, \ and\ \bibinfo {author} {\bibfnamefont {Y.}~\bibnamefont {Chen}},\ }\bibfield  {title} {\enquote {\bibinfo {title} {Asymmetry of the atomic-level stress tensor in homogeneous and inhomogeneous materials},}\ }\href@noop {} {\bibfield  {journal} {\bibinfo  {journal} {Proc. R. Soc. A.}\ }\textbf {\bibinfo {volume} {474}},\ \bibinfo {pages} {20180155} (\bibinfo {year} {2018})}\BibitemShut {NoStop}%
\end{thebibliography}

\end{document}